\documentclass[pra,aps,twocolumn,superscriptaddress, longbibliography, notitlepage]{revtex4-1}
\usepackage[colorlinks=true, citecolor=blue, urlcolor=blue, linkcolor=blue]{hyperref}
\usepackage{graphicx}
\usepackage{dcolumn}
\usepackage{bm}
\usepackage{amsmath,amsfonts}
\usepackage[english]{babel}
\usepackage{amsthm}
\usepackage{hyperref}
\usepackage{xcolor}
\theoremstyle{plain}

\newtheorem{theorem}{Theorem}

\newtheorem{lemma}{Lemma}

\begin{document}
\title{Unbounded sequential multipartite nonlocality via  violation of Mermin inequality}

\author{Bang-Zhu Shen}
\affiliation{School of Mathematics, South China University of Technology, GuangZhou 510640, China}
\author{Mao-Sheng Li}
\email{li.maosheng.math@gmail.com}
\affiliation{School  of Mathematics, South China University of Technology, GuangZhou 510640, China}

\begin{abstract}
Quantum nonlocality is a significant feature in  quantum information theory, prompting recent investigations into the potential reuse of post-measurement states to uncover nonlocality among sequentially measuring observers. While prior studies primarily focused on bipartite or tripartite systems  and observers with  one chain, such as multiple Bobs with a single Alice or multiple Charlies with a single Alice and Bob, our work extends beyond this framework. We explore sequential nonlocality in systems comprising more parties and observer chains. Our findings reveal that in $n$-partite systems, regardless of whether it is a single-chain or double-chain scenario, there exist unbounded sequential observers capable of detecting nonlocality through violations of the Mermin inequality. In contrast to the conjecture that sequential Bell nonlocality cannot manifest with multiple Alices and Bobs  in bipartite systems  (i.e., the double-chain setting)[\href{https://doi.org/10.1103/PhysRevA.104.L060201}{Phys. Rev. A \textbf{104}, L060201 (2021)}], our results suggest that increasing the number of  subsystems may enable more observer chains to detect nonlocality alongside single observers. Our study advances research on sequential  nonlocality, providing valuable insights into its detection across diverse scenarios.

\end{abstract}
\maketitle
\section{INTRODUCTION}
Quantum nonlocality stands as a pivotal element of quantum information, often validated through the violation of Bell inequalities \cite{1}. It constitutes a fundamental aspect of quantum mechanics and a cornerstone resource in quantum information science, boasting a lot of applications, such as proposing nonlocality distillation protocols, making communication complexity trivial \cite{2,3} and providing valuable insights into the possibility of device-independent scenarios for quantum key distribution (QKD) \cite{4,5}.

Recently, researchers have been exploring the possibility of violating nonlocality sequentially using a single pair of entangled qubits. In Ref. \cite{6}, the authors considered the sequential Clauser-Horne-Shimony-Holt (CHSH) scenario and demonstrated that multiple Bobs can achieve expected violation of CHSH inequality with a single Alice. Subsequently, in Ref. \cite{7,8}, the authors demonstrated that multiple Bobs can achieve expected violation of CHSH inequality with a single Alice in high dimensions. Additionally, several
studies have explored the recyclability of nonlocal correlations \cite{9,10,11,12,13,14,15,Hu22,Kumari23}, provided evidence that the weak measurement strategy can be used to achieve an arbitrarily long sequence of Alice-Bob pairs for any pure entangled state \cite{16,17}. 

Tripartite settings have also got attention in this domain. In Ref. \cite{7}, it was demonstrated that, based on the Svetlichny inequality \cite{18}, at most two Charlies sharing genuine nonlocality with Alice and Bob in their measurement. Meanwhile, researchers in Ref. \cite{19} investigated tripartite quantum nonlocality using various inequalities such as the Mermin inequality and nonsignal (NS) inequality, along with different initial states like the W state and Greenberger-Horne-Zeilinger (GHZ) state. They found that in the case of starting with a GHZ state, tripartite nonlocality can be demonstrated by any number of Charlies using the Mermin inequality and NS inequality in their measurement. However, when starting with a W state, at most two Charlies can detect genuinely tripartite nonlocality via the violation of the Mermin inequality in their measurement. Furthermore, experimental studies \cite{20,21} have shown violations of Mermin inequality in multiple qubits, such as in a five-qubit quantum computer and a 53-qubit system, providing empirical support for these theoretical findings. \\
\indent Consequently, it is pertinent to investigate whether such properties persist as the number of parties involved increases. In this study, we aim to provide affirmative responses regarding single-chain and double-chain scenarios where unbounded multipartite nonlocality can be established through the violation of the Mermin inequality.
\\
\indent The structure of this paper is as follows. In Sec. \ref{s1}, we review the concept of standard nonlocality and the Mermin inequality, which is instrumental in detecting such nonlocality. And we have obtained the general formula for the coefficients of the Mermin polynomials. In Sec. \ref{s2}, we initially present the single-chain scenario. By employing the Mermin inequality, we demonstrate the existence of unbounded sequential tripartite nonlocality for an initially shared W state, as well as unbounded sequential $n$-partite $(n\geq3)$ nonlocality for an initially shared GHZ state. In Sec. \ref{s5}, we proceed to introduce the double-chain scenario. Utilizing the Mermin inequality once again, we reveal the presence of unbounded sequential $n$-partite $(n\geq4)$ nonlocality for an initially shared GHZ state. Finally, in Sec. \ref{s8}, we summarize our findings, draw conclusions, and suggest topics that merit further investigation.

\section{MERMIN POLYNOMIALS AND THE DETECTION OF STANDARD $n$-PARTITE NONLOCALITY VIA MERMIN INEQUALITIES}\label{s1}

Firstly, we consider $n$-parties separated parties, denoted as $A^{(1)}$, $A^{(2)}$, $\cdots$, $A^{(n)}$, sharing an 
$n$-partite physical systems $\rho$. In the simplest scenario, each party $A^{(i)}$
  has two measurements $\{M_{0|x_i}^{(i)}, M_{1|x_i}^{(i)} \}_{x_i\in\mathbb{Z}_2}$ (here and the following $\mathbb{Z}_2:=\{0,1\}$) where each $M_{a_i|x_i}\geq 0 $ (i.e., a positive semidefinite operator) and  $M_{0|x_i}^{(i)}+M_{1|x_i}^{(i)}=\mathbb{I}$.   And we define the observable 
  $M_{x_i}^{(i)}$ corresponding the  measurement $ \{M_{0|x_i}^{(i)}, M_{1|x_i}^{(i)} \}$  by setting 
  $$M_{x_i}^{(i)}:=\sum_{a_i\in \mathbb{Z}_2}  (-1)^{a_i} M_{a_i|x_i}^{(i)} =M_{0|x_i}^{(i)} -M_{1|x_i}^{(i)}.$$ So the outcome is $(-1)^{a_i}$   but will be labeled as $a_i$. Note that $-\mathbb{I}\leq M_{x_i}^{(i)}\leq \mathbb{I}$ and  the measurement $ \{M_{0|x_i}^{(i)}, M_{1|x_i}^{(i)} \}$  can be also uniquely identified by its observable $M_{x_i}^{(i)}$. In fact,   provided the condition  $-\mathbb{I}\leq M_{x_i}^{(i)}\leq \mathbb{I}$, the following two elements form a measurement with observable  $M_{x_i}^{(i)}$ 
  $$M_{0|x_i}^{(i)}=\frac{\mathbb{I}+M_{x_i}^{(i)} }{2},  \  \text{ and } \ \ M_{1|x_i}^{(i)}=\frac{\mathbb{I}-M_{x_i}^{(i)} }{2}.$$ The spatially separated parties $A^{(1)}$, $A^{(2)}$, $\cdots$, $A^{(n)}$ randomly choose their two observables and take the corresponding  measurements $\{M_{0|x_i}^{(i)}, M_{1|x_i}^{(i)} \}$, yielding outcomes $a_{i}\in\{0,1\}$. 
  We define the joint outcome probabilities as $P(a_{1}{\cdots} a_{n}|x_{1}{\cdots} x_{n})$. If these probability correlations can be expressed as
\begin{equation*}
 P(a_{1}{\cdots} a_{n}|x_{1}{\cdots} x_{n})=\sum\limits_{\lambda}q_{\lambda}\prod\limits_{i}P_{\lambda}(a_{i}|x_{i})
\end{equation*}
with $q_{\lambda} \in [0,1]$ and $\sum\limits_{\lambda}q_{\lambda}=1$. Then they are called fully local. If they are not fully local, we say $P(a_{1}{\cdots} a_{n}|x_{1}{\cdots} x_{n})$ exhibits \emph{standard $n$-partite nonlocality} \cite{19,24}.

  Secondly, referring to \cite{22,23}, 
  the multipartite Mermin polynomials $M_n$ can be calculated as follows 
\begin{equation}\label{m1}
   \begin{array}{l}
           \displaystyle M_n=\frac{M_{n-1}(M_{0}^{(n)}+M_{1}^{(n)})}{2}+\frac{M'_{n-1}(M_{0}^{(n)}-M_{1}^{(n)})}{2},\\[2mm]
     \displaystyle  M'_n=\frac{M_{n-1}(M_{1}^{(n)}-M_{0}^{(n)})}{2}+\frac{M'_{n-1}(M_{0}^{(n)}+M_{1}^{(n)})}{2}.
  \end{array}
\end{equation}
with $M_1=M_{0}^{(1)}$, $M'_1=M_{1}^{(1)}$.
Specifically, standard nonlocality can be detected by violation of the Mermin inequality \cite{21,22,23,24},
 i.e., \begin{equation}\label{mk}
    \langle M_n\rangle:=\mathrm{Tr}[ \rho  M_n] > 1.\end{equation}
Denote $\mathbf{v}=(v_1,v_2,\cdots, v_n)$ as a vector in $\mathbb{Z}_2^n$, and we define $|\mathbf{v}|$:=$\sum_{i=1}^nv_i$. It is found that $M_n$ can be represented by the following expression:
\begin{equation}\label{m5}
    M_n=\sum_{\mathbf{v}\in \mathbb{Z}_2^n} c_{\mathbf{v}} \prod_{i=1}^n M_{v_i}^{(i)}
\end{equation}
where $c_{\mathbf{v}}\in \mathbb{R}.$
For instance, in a tripartite setting, the coefficients of $M_3$ are specified as follows
$ c_{(1,0,0)}$=$c_{(0,1,0)}$=$c_{(0,0,1)}=\frac{1}{2}$ ,  $c_{(1,1,1)}=-\frac{1}{2} $, and all other coefficients are zero. Therefore, 
$$\begin{array}{rcl}

  M_3&=&\frac{1}{2} M_1^{(1)}M_0^{(2)}M_0^{(3)}+\frac{1}{2} M_0^{(1)}M_1^{(2)}M_0^{(3)} 
 \\[2mm]
 && +\frac{1}{2} M_0^{(1)}M_0^{(2)}M_1^{(3)}-\frac{1}{2} M_1^{(1)}M_1^{(2)}M_1^{(3)}.
\end{array}$$
Furthermore, when considering $n$-partite systems, the coefficients $c_{\mathbf{v}}$ of $M_{n}$ in Eq.\eqref{m5}  can be expressed  by (see Appendix \ref{ap1}) \begin{equation}\label{eq:c_Expression}
    c_{\mathbf{v}}=\frac{1}{2} \left((\frac{1}{\sqrt{2}}\lambda_2)^{n-1} \lambda_1^{2|\mathbf{v}|}+(\frac{1}{\sqrt{2}}\lambda_1)^{n-1} \lambda_2^{2|\mathbf{v}|}\right),
\end{equation} 
where  $\lambda_1=e^{\pi  i/4}$ and  $\lambda_2=e^{-\pi  i/4}.$

In this work, we introduce the Mermin values $\mathbf{I}_{n}$:=$\langle M_n\rangle$. Based on the standard multipartite nonlocality criterion given by Eq. (\ref{mk}), the standard nonlocality can be detected when \begin{equation}\label{m}
    \mathbf{I}_{n} > 1,
\end{equation}
 for $n$-partite systems. In particular, the $\langle M_n\rangle $ depends on the state $\rho$ and the observables $\{M_{x_i}^{(i)}\}_{x_i\in\mathbb{Z}_2}, i=1,2,\cdots, n.$ 
\section{SHARING OF NONLOCALITY VIA MERMIN INEQUALITY IN THE SINGLE-CHAIN SCENARIO}\label{s2}
\begin{figure}[ptb]
    \centering
    \includegraphics[width=0.5\textwidth]{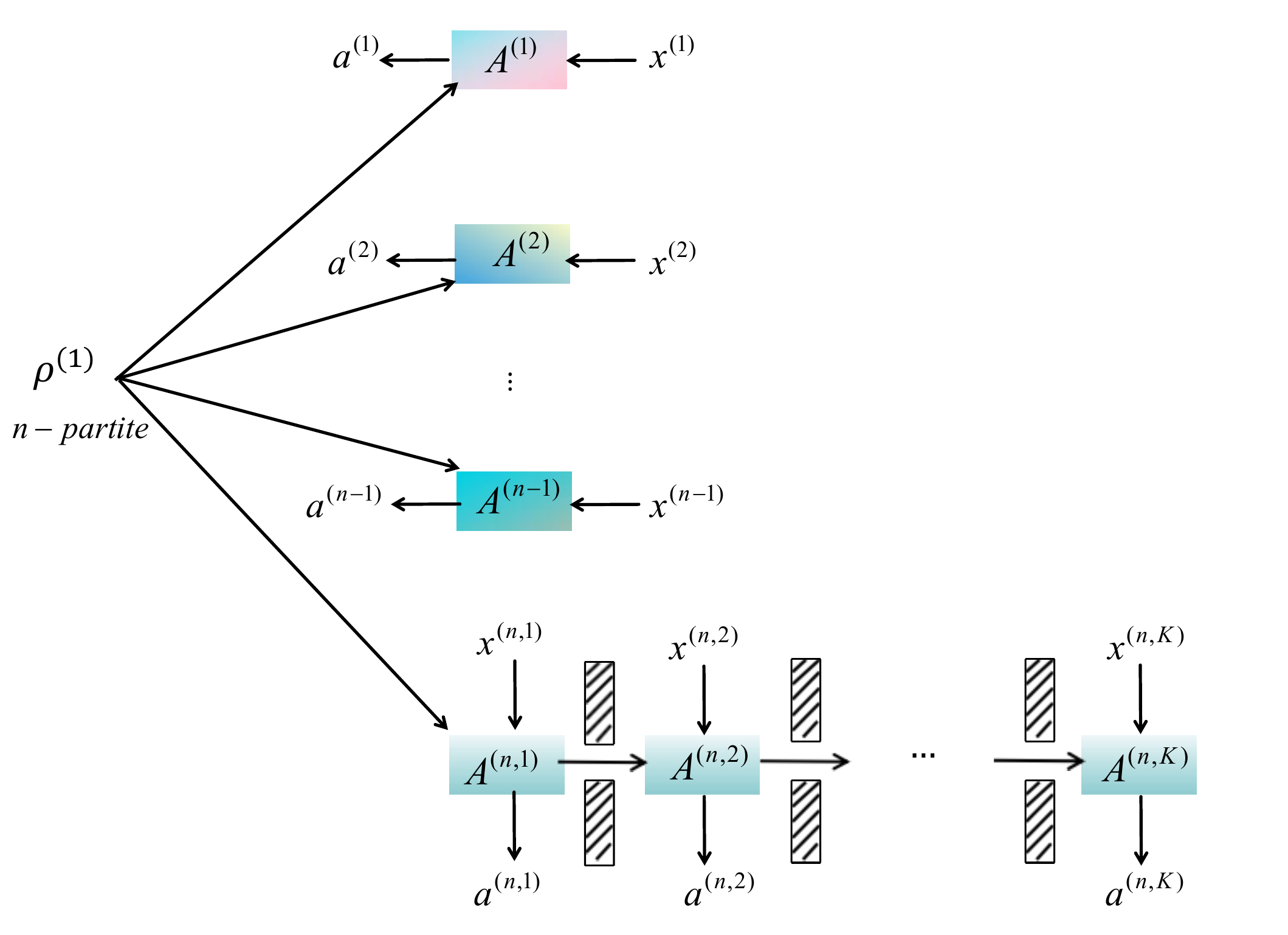}
    \caption{Sharing the $n$-partite nonlocality: an entangled quantum state $\rho^{(1)}$ is initially distributed among $A^{(1)}$, $A^{(2)}$, $\cdots$, $A^{(n-1)}$, $A^{(n,1)}$. After $A^{(n,1)}$ performs his randomly selected measurement and records the outcomes, he passes the postmeasurement quantum state to $A^{(n,2)}$. This sequence of actions is reiterated along the single chain, ultimately reaching $A^{(n,K)}$. }
    \label{f1}
\end{figure}

In this section, for any $K\in \mathbb{N}$, we introduce the single-chain scenario where $A^{(1)}$, $A^{(2)}$, $\cdots$, $A^{(n-1)}$ attempt to share the nonlocal correlation of an entangled pure state $\rho^{(1)}$ with  a chain of $K$ independent  agents  $\{A^{(n,1)}, \cdots, A^{(n,K)}\}$ (see Figure. \ref{f1}).
Initially, let $\rho^{(1)}$ be an $n$-partite entangled pure state shared by $A^{(1)}$, $A^{(2)}$, $\cdots$, $A^{(n-1)}$, and $A^{(n,1)}$. After $A^{(n,1)}$ performs his randomly selected measurement and records the outcomes, he passes the postmeasurement quantum state to $A^{(n,2)}$. We denote the binary input and output of $A^{(i)}$ ($A^{(n,j)}$) as $x^{(i)}$ ($x^{(n,j)}$) and $a^{(i)}$ ($a^{(n,j)}$) respectively, where $i\in\{1, 2, \cdots, n-1\}, j \in\{1, 2, \cdots, K\}$. Assuming $A^{(n,1)}$ performs the measurement based on the input $x^{(n,1)}$ and obtains the outcome $a^{(n,1)}$. The postmeasurement state can be described by the L\"{u}ders rule as follows:
$$\begin{array}{rcl}\rho^{(2)}&=&\frac{1}{2}\sum\limits_{a^{(n,1)},x^{(n,1)}}(\mathbb{I}^{\otimes(n-1)}\otimes\sqrt{M^{(n,1)}_{a^{(n,1)}|x^{(n,1)}}})\rho^{(1)}\\[2mm]&&(\mathbb{I}^{\otimes(n-1)}\otimes\sqrt{M^{(n,1)}_{a^{(n,1)}|x^{(n,1)}}}),
\end{array}$$
where $M^{(n,1)}_{a^{(n,1)}|x^{(n,1)}}$ represents the POVM effect corresponding to the outcome $a^{(n,1)}$ of $A^{(n,1)}$'s obersevable $M_{ x^{(n,1)}}^{(n,1)}$. By repeating this process up to $A^{(n,K)}$, we can get the state
$$\begin{array}{rcl}
    \rho^{(k)}&=&\frac{1}{2}\sum\limits_{a^{(n,k-1)},x^{(n,k-1)}}\left(\mathbb{I}^{\otimes(n-1)}\otimes\sqrt{M^{(n,k-1)}_{a^{(n,k-1)}|x^{(n,k-1)}}}\right)\\[3mm]&&\rho^{(k-1)}\left(\mathbb{I}^{\otimes(n-1)}\otimes\sqrt{M^{(n,k-1)}_{a^{(n,k-1)}|x^{(n,k-1)}}}\right),
\end{array}
$$where $ 1 \leq k\leq K$.

 We firstly consider the tripartite nonlocality for an initially shared $W$ state. Subsequently, we consider  $n$-partite($n\geq3$) nonlocality  for an initially shared GHZ state.

\subsection{TRIPARTITE NONLOCALITY STARTING FROM   $W$  STATE}\label{s3}
In this subsection,  we explore tripartite nonlocality involving an arbitrary number of Charlies, a single Alice, and a single Bob. 
Referring to \cite{19}, it was demonstrated that, with a single Alice and Bob, any number of Charlies can detect standard nonlocality by violating the Mermin inequality when they initially share a generalized GHZ state. However,  in their measurement, it was shown that at most two Charlies can detect standard nonlocality for an initially shared $W$  state. Here we establish a more robust result: an arbitrary number of Charlies can violate the Mermin inequality with a single Alice and Bob when starting from an initially shared $W$  state.

To detect the nonlocality, it is essential to examine the Mermin value $\mathbf{I}^{(k)}_{3}$, which involves  Alice, Bob and Charlie$^{(k)}$. 
Therefore, We need to define a proper measurement strategy for Alice, Bob, Charlie$^{(k)}$.   In this measurement strategy, Alice's observables are defined by
\begin{eqnarray*}\label{eq:A1}
  M^{(1)}_0=\sin(\theta)\sigma_{1}+\cos(\theta)\sigma_{3}, \ \
  M^{(1)}_1=   \sin(\theta)\sigma_{1} -\cos(\theta)\sigma_{3}.
\end{eqnarray*}
Bob's observables are defined by
\begin{eqnarray*}\label{eq:A1}
  M^{(2)}_0=\sin(\theta)\sigma_{1}+\cos(\theta)\sigma_{3}, \ \
  M^{(2)}_1=   \sin(\theta)\sigma_{1} -\cos(\theta)\sigma_{3}.
\end{eqnarray*}
 Charlie$^{(k)}$'s ($1 \leq k \leq K$) observables are defined by
\begin{eqnarray*}\label{eq:B1}
  M^{(3,k)}_0=\sigma_{3}, \ \
  M^{(3,k)}_1= \gamma_k(\theta)\sigma_{1},  \ \  (0<\gamma_k(\theta)<1).
\end{eqnarray*}

Set  $\left|\psi\right\rangle=\frac{1}{\sqrt{3}}(\left|100\right\rangle+\left|010\right\rangle+\left|001\right\rangle)$ and  let $\rho^{(1)}=\left|\psi\right\rangle\left\langle\psi\right|$    be the initial state shared by Alice, Bob, and   Charlie$^{(1)}$.  After, the first $(k-1)$ Charlies taking their measurements,  the state $\rho^{(k)}$ shared by  Alice, Bob, and   Charlie$^{(k)}$ is given by the  recursive relation 
$$\rho^{(k)}=\frac{1}{2}\sum\limits_{c,z}(\mathbb{I}^{\otimes 2}\otimes\sqrt{M^{(3,k-1)}_{c|z}})\rho^{(k-1)}(\mathbb{I}^{\otimes 2}\otimes\sqrt{M^{(3,k-1)}_{c|z}}).$$
Given these
measurements and the initial state, we can calculate the Mermin value $\mathbf{I}^{(k)}_{3}(\theta)$  with respect to the state $\rho^{(k)}$ and   the observables $\{M^{(1)}_x\}_{x\in\mathbb{Z}_2}, $ $\{M^{(2)}_y\}_{y\in\mathbb{Z}_2}, $ $\{M^{(3,k)}_z\}_{z\in\mathbb{Z}_2}$ as follows
\begin{eqnarray}\label{xm}
    \frac{P_k(\theta)(2\cos^2(\theta)+\frac{4\sin^2{(\theta)}}{3})}{2^{k}}+\frac{8\sin(\theta)\cos(\theta)\gamma_k(\theta)}{3\cdot2^{k}}.
\end{eqnarray}
Here  $P_{k}(\theta)=\prod\limits_{j=1}^{k-1}(1+\sqrt{1-\gamma_{j}^{2}(\theta)})$ (see Appendix \ref{ap2} for the detailed calculation).

To ensure $\mathbf{I}^{(k)}_{3}(\theta)>1$, we solve for the appropriate $\gamma_k(\theta)$ using Eq. (\ref{xm})), which leads to the inequality:
\begin{eqnarray}\label{}
\gamma_k(\theta)>\frac{2^k-P_k(\theta)(2-\frac{2\sin^2{(\theta)}}{3})}{\frac{8\sin(\theta)\cos(\theta)}{3}}.
\end{eqnarray}
To achieve it, for $\forall\epsilon>0$, we propose a specific sequence $\{\gamma_k(\theta)\}$, defined as:
\begin{equation}\label{}
  \gamma_{k}(\theta)=\left\{
                       \begin{array}{ll}
                         (1+\epsilon)(\frac{\tan(\theta)}{4}), & \hbox{$k=1$;} \\
                     (1+\epsilon)(\frac{2^k-P_k(\theta)(2-\frac{2\sin^2{(\theta)}}{3})}{\frac{8\sin(\theta)\cos(\theta)}{3}}) , & \hbox{$0\leq\gamma_{k-1}(\theta)\leq1$ ;} \\
                         \infty, & \hbox{otherwise.}
                       \end{array}
                     \right.
\end{equation}
We now present the following theorem, which supports our initial claim that an arbitrary number of Charlies can violate the Mermin inequality in conjunction with a single Alice and Bob.
\begin{theorem}\label{thm:Main1}
    For $\forall$ $K\in\mathbb{N}$, there exists a sequence $\{\gamma_{k}(\theta)\}_{k=1}^{K}$, $\theta_{K} \in(0,1)$ and $\theta \in(0,\theta_{K})$ such that $\mathbf{I}^{(k)}_{3}(\theta)>1$ and $\gamma_k(\theta) \in(0,1)$ for $k\in\{1,2,\cdots,K\}$.
\end{theorem}
This proof is given in the Appendix \ref{ap3}. Theorem  \ref{thm:Main1} demonstrates that unbounded sequential tripartite nonlocality can be detected for the initially shared W state in the single-chain setting, which is a significant finding in the study of quantum correlations.

\subsection{$n$-PARTITE NONLOCALITY STARTING FROM  GHZ  STATE}\label{s4}
In this subsection, we consider $n$-partite ($n\geq 3$) nonlocality involving $A^{(1)}$, $A^{(2)}$, $\cdots$, $A^{(n-1)}$, and $A^{(n,k)}$ $(1\leq k \leq K)$. To detect the nonlocality, it is essential to examine the Mermin value $\mathbf{I}^{(k)}_{n}$. Therefore, we need to define a proper measurement strategy by the following measurement for $A^{(1)}$, $A^{(2)}$, $\cdots$, $A^{(n-1)}$, and $ A^{(n,k)}$. Next, we give the corresponding measurement strategies in two cases. If $n\equiv 0$ or $1$ or $2 \mod 4,$ we take the following measurement strategy, $A^{(j)}$'s observables are defined by
\begin{eqnarray}\label{eq:A1}
 M^{(j)}_0 =\sigma_{1}, \ \
  M^{(j)}_1 = \sigma_{2}.
\end{eqnarray}
where $j=1,2,\cdots,n-2$.
$A^{(n-1)}$'s observables are defined by
\begin{eqnarray}\label{eq:A1}
 M^{(n-1)}_0 =\theta\sigma_{1}, \ \
  M^{(n-1)}_1 = \theta\sigma_{2}.
\end{eqnarray}
$A^{(n,k)}$'s  $(1\leq k\leq K)$ observables are defined by
\begin{eqnarray}\label{eq:A1}
 M^{(n,k)}_0 =\sigma_{1}, \ \
  M^{(n,k)}_1 = \gamma_{k}(\theta)\sigma_{2},  \ \  (0<\gamma_{k}(\theta)<1).
\end{eqnarray}
If $n\equiv 3 \mod 4,$ we change the $A^{(n-1)}$'s $(1\leq k \leq K)$ observables   as follows  
\begin{eqnarray}\label{eq:casemod3}
 M^{(n-1)}_0 =-\theta\sigma_{2}, \ \
  M^{(n-1)}_1 = \theta\sigma_{1}.
\end{eqnarray}
while the other are the same.

Set $\left|\mathrm{GHZ}_n\right\rangle=\frac{1}{\sqrt{2}}(\left|00\cdots0\right\rangle+\left|11\cdots1\right\rangle)$, and let $\rho^{(1)}=\left|\mathrm{GHZ}_n\right\rangle\left\langle\mathrm{GHZ}_n\right|$    be the initial state shared by $A^{(1)}$, $A^{(2)}$, $\cdots$, $A^{(n-1)}$, and $A^{(n,1)}$.  
Given these
measurements and the initial state, we can calculate the Mermin value $\mathbf{I}^{(k)}_{n}(\theta)$  with respect to the state $\rho^{(k)}$ and the observables $\{M_{x_i}^{(i)}\}_{x_i\in\mathbb{Z}_2}$ for $i=1,2,\cdots, n-1$ and  $\{M_{x_n}^{(n,k)}\}_{x_n\in\mathbb{Z}_2}$ as follows (see Appendix \ref{ap4} for the detailed calculation) 
\begin{equation}\label{eq:Merminn1} \mathbf{I}^{(k)}_{n}(\theta) =\displaystyle N_n\frac{\theta\gamma_k(\theta)+\theta  P_k(\theta)}{2^{k-1}}, \end{equation}
where $P_{k}(\theta)=\prod\limits_{j=1}^{k-1}(1+\sqrt{1-\gamma_{j}^{2}(\theta)})$ and  \begin{equation}\label{eq:Constants1}
    N_n=\begin{cases}
        (\sqrt{2})^{n-3}(-1)^ {\lfloor \frac{n}{4}\rfloor},  &\text{if } n \equiv 1 \text{ or }   3  \mod 4,\\[2mm](\sqrt{2})^{n-4}(-1)^ {\lfloor \frac{n}{4}\rfloor},  &\text{if } n \equiv 0 \text{ or }  2  \mod 4.
    \end{cases}
\end{equation}

For $\forall K\in\mathbb{N}$, $k\in\{1, 2, \cdots, K\}$, our goal is to select appropriate values for $\theta$ and find a sequence $\{\gamma_{k}(\theta)\}_{k=1}^{K}$ such that $\mathbf{I}^{(k)}_{n}(\theta)>1$. Hence, we propose the following lemma to solve this problem.
\begin{lemma}\label{Lemma} Let $N$  be a real number with $|N|\geq 1$.
    For  any $  K\in\mathbb{N}$, there exist some $\theta\in(-1,1)$ and a sequence $\{\gamma_{k}(\theta)\}_{k=1}^{K}$  such that     $\gamma_k(\theta)\in(0,1)$ and $$f_k(\theta)=N\frac{\theta \gamma_k(\theta)+\theta P_k(\theta)}{2^{k-1}}>1$$  for each $1\leq k\leq K$, where $P_{k}(\theta)=\prod\limits_{j=1}^{k-1}(1+\sqrt{1-\gamma_{j}^{2}(\theta)}).$ 
\end{lemma}

The proof of Lemma \ref{Lemma} is given in Appendix \ref{ap5}.  

From Eq.  \eqref{eq:Constants1}, it's easy to check that $|N_n|\geq1$  whenever $n\geq 3$. Then, using Lemma \ref{Lemma},  we can deduce the following theorem.

\begin{theorem}\label{thm:Main2}
    For $\forall$ $n\in\mathbb{N}$, there exist some $\theta\in(-1,1)$ and a sequence $\{\gamma_{k}(\theta)\}_{k=1}^{n}$  such that $\gamma_k(\theta)\in(0,1)$ and  $\mathbf{I}^{(k)}_{n}(\theta) >1$  for $k\in\{1,2,\cdots,K\}$. That is,  we can detect unbounded sequential 
$n$-partite standard nonlocality  starting from  $\mathrm{GHZ}$ state in the single-chain setting.
\end{theorem}

\section{SHARING OF $n$-PARTITE NONLOCALITY VIA MERMIN INEQUALITY IN THE DOUBLE-CHAIN SCENARIO}\label{s5}
\begin{figure}[ptb]
    \centering
    \includegraphics[width=0.5\textwidth]{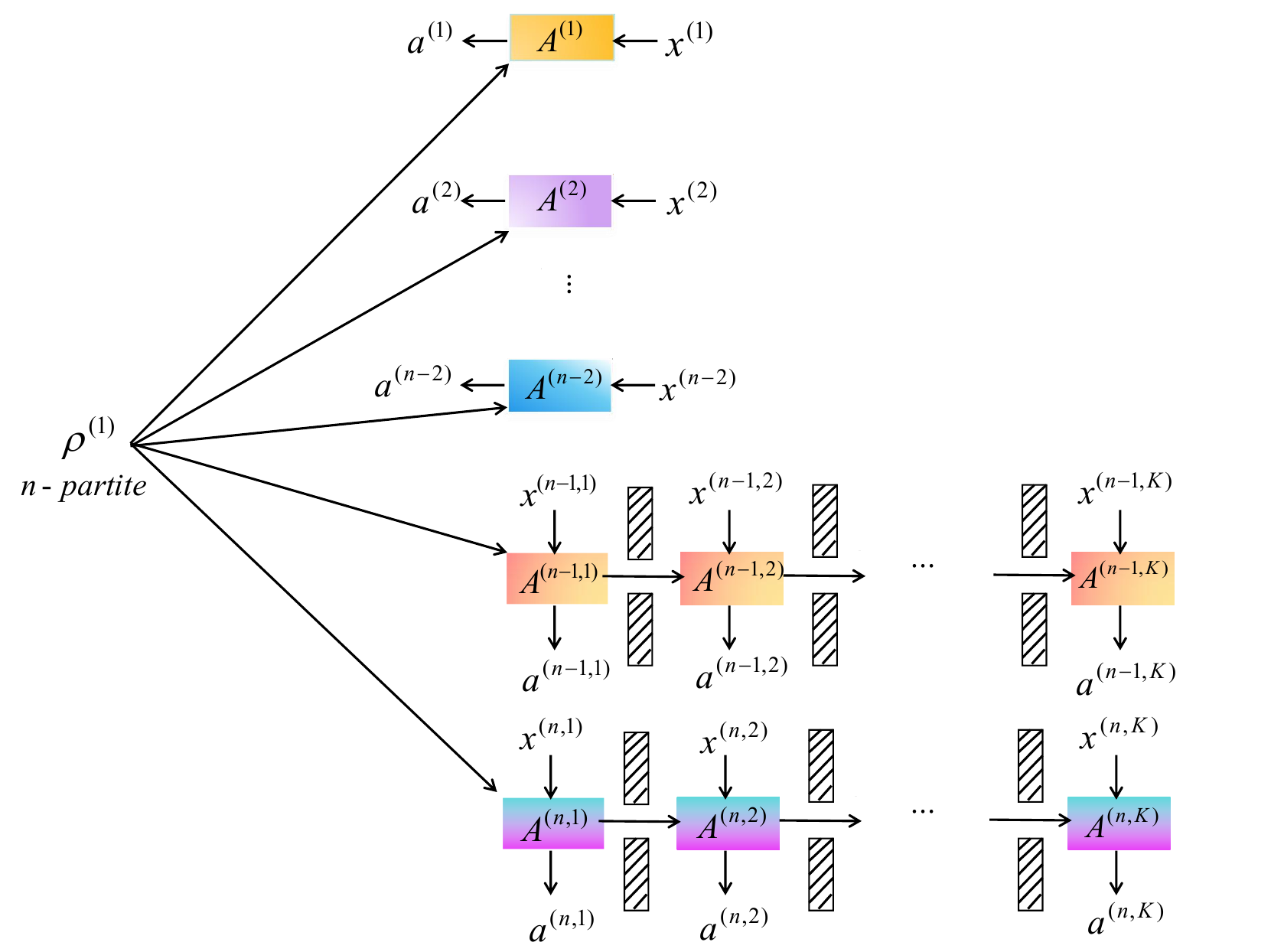}
    \caption{Sharing the $n$-partite nonlocality: a quantum state $\rho^{(1)}$ is initially distributed among $A^{(1)}$, $A^{(2)}$, $\cdots$, $A^{(n-1)}$, $A^{(n-1,1)},$ $A^{(n,1)}$. Subsequently, $A^{(n-1,1)}$ and $A^{(n,1)}$ perform their randomly selected measurement, record the outcomes and passes the postmeasurement quantum state to $A^{(n-1,2)}$ and $A^{(n,2)}$ respectively. This sequence of actions is reiterated along two chains respectively, ultimately reaching $A^{(n-1,K)}$ and $A^{(n,K)}$.}
    \label{f2}
\end{figure}

In this section, for any $K\in \mathbb{N}$, we introduce the double-chain scenario where $A^{(1)}$, $A^{(2)}$, $\cdots$, $A^{(n-2)}$ attempt to share the nonlocal correlation of an entangled pure state $\rho^{(1)}$ with  two chains of  independent agents and each chain has $K$  parties $\{A^{(n-1,1)}, \cdots, A^{(n-1,K)}\}$ and  $\{A^{(n,1)}, \cdots, A^{(n,K)}\}$  respectively (see Figure. \ref{f2}).

To begin with, let $\rho^{(1)}$ be an $n$-partite entangled pure state shared by $A^{(1)}$, $A^{(2)}$, $\cdots$, $A^{(n-2)}$, $A^{(n-1,1)}$ and $A^{(n,1)}$. After $A^{(n-1,1)}$ and $A^{(n,1)}$ perform their randomly selected measurement and record the outcomes, they pass the postmeasurement quantum state to $A^{(n-1,2)}$ and $A^{(n,2)}$, respectively. We denote the binary input and output of $A^{(i)}$ ($A^{(n-1,j)}$, $A^{(n,j)}$) as $x^{(i)}$ ($x^{(n-1,j)}$, $x^{(n,j)}$) and $a^{(i)}$ ($a^{(n-1,j)}$, $a^{(n,j)}$) respectively, where $i\in\{1, 2, \cdots, n-2\}$ and $j \in\{1, 2, \cdots, K\}$. Assuming $A^{(n-1,1)}$ and $A^{(n,1)}$ perform measurements based on the inputs $x^{(n-1,1)}$ and $x^{(n,1)}$, obtain the outcomes $a^{(n-1,1)}$ and $a^{(n,1)}$. The postmeasurement state is described by the L\"{u}ders rule as follows:
$$\begin{array}{rcl}
\rho^{(2)}&=&\frac{1}{4}\sum\Big (\mathbb{I}^{\otimes(n-2)}\otimes\sqrt{M^{(n-1,1)}_{a^{(n-1,1)}|x^{(n-1,1)}}}\\[3mm]&&\otimes\sqrt{M^{(n,1)}_{a^{(n,1)}|x^{(n,1)}}}\Big)\rho^{(1)}\Big(\mathbb{I}^{\otimes(n-2)}\\[3mm]&&\otimes\sqrt{M^{(n-1,1)}_{a^{(n-1,1)}|x^{(n-1,1)}}}\otimes\sqrt{M^{(n,1)}_{a^{(n,1)}|x^{(n,1)}}}\Big).
\end{array}$$
The summation involves the terms $a^{(n-1,1)}$, $a^{(n,1)}$, $x^{(n-1,1)}$ and $x^{(n,1)}$, each of which can take a value of either 0 or 1. At the same time, $M^{(n-1,1)}_{a^{(n-1,1)}|x^{(n-1,1)}}$ and $M^{(n,1)}_{a^{(n,1)}|x^{(n,1)}}$ represent the POVM effects corresponding to the outcomes $a^{(n-1,1)}$ and $a^{(n,1)}$ of $A^{(n-1,1)}$'s and $A^{(n,1)}$'s measurements, respectively. By repeating this process up to $A^{(n-1,K)}$ and $A^{(n,K)}$, we obtain the state

$$\begin{array}{rcl}
\rho^{(k)}&=&\frac{1}{4}\sum\Big(\mathbb{I}^{\otimes(n-2)}\otimes\sqrt{M^{(n-1,k-1)}_{a^{(n-1,k-1)}|x^{(n-1,k-1)}}}\\[3mm]&&\otimes\sqrt{M^{(n,k-1)}_{a^{(n,k-1)}|x^{(n,k-1)}}}\Big)\rho^{(1)}\Big(\mathbb{I}^{\otimes(n-2)}\\[3mm]&&\otimes\sqrt{M^{(n-1,k-1)}_{a^{(n-1,k-1)}|x^{(n-1,k-1)}}}\otimes\sqrt{M^{(n,k-1)}_{a^{(n,k-1)}|x^{(n,k-1)}}}\Big).
\end{array}$$
The summation involves the terms $a^{(n-1,k-1)}$, $a^{(n,k-1)}$, $x^{(n-1,k-1)}$, $x^{(n,k-1)}$ as above with $1\leq k \leq K$.


Then, we consider $n$-partite ($n\geq 4$) nonlocality involving $A^{(1)}$, $A^{(2)}$, $\cdots$, $A^{(n-2)}$, $A^{(n-1,k)}$, and $A^{(n,k)}$.
Following standard procedures, we aim to calculate the Mermin value ${\mathbf{I}^{(k)}_{n}}$ among $A^{(1)}$, $A^{(2)}$, $\cdots$, $A^{(n-2)}$, $A^{(n-1,k)}$, and $A^{(n,k)}$.

Therefore, we define a detailed measurement strategy for $A^{(1)}$, $A^{(2)}$, $\cdots$, $A^{(n-2)}$, $A^{(n-1,k)}$, and $A^{(n,k)}$. We also consider two situations like single-chain setting. If $n\equiv 0$ or $1$ or $3 \mod 4,$ we take the following measurement strategy,  $A^{(1)}$'s observables are defined by
\begin{eqnarray}\label{eq:A1}
 M^{(1)}_0 =\theta\sigma_{1}, \ \
  M^{(1)}_1 = \theta\sigma_{2}.
\end{eqnarray}
$A^{(j)}$'s observables are defined by
\begin{eqnarray}\label{eq:A1}
 M^{(j)}_0 =\sigma_{1}, \ \
  M^{(j)}_1 = \sigma_{2}.
\end{eqnarray}
where $j=2,\cdots,n-2$.
 $A^{(n-1,k)}$'s observables $(1\leq k \leq K)$ are defined by
\begin{eqnarray}\label{eq:A1}
 M^{(n-1,k)}_0 =\sigma_{2}, \ \
  M^{(n-1,k)}_1 = \sigma_{2}.
\end{eqnarray}
A$^{(n,k)}$'s $(1\leq k \leq K)$ observables are defined by
\begin{eqnarray}\label{eq:A1}
 M^{(n,k)}_0 =\sigma_{1}, \ \
  M^{(n,k)}_1 =  {\gamma}_{k}(\theta)\sigma_{2},\ \  (0< {\gamma}_{k}(\theta)<1).
\end{eqnarray} And if $n\equiv 2 \mod 4,$ we change the $A^{(n-1,k)}$'s $(1\leq k \leq K)$ observables   as follows  
\begin{eqnarray}\label{gm9}
 M^{(n-1,k)}_0 =-\sigma_{2}, \ \
  M^{(n-1,k)}_1 = \sigma_{2}.
\end{eqnarray}  
while the other are the same.

Set $\left|\mathrm{GHZ}_n\right\rangle=\frac{1}{\sqrt{2}}(\left|00\cdots0\right\rangle+\left|11\cdots1\right\rangle)$, and let $\rho^{(1)}=\left|\mathrm{GHZ}_n\right\rangle\left\langle\mathrm{GHZ}_n\right|$    be the initial state shared  by $A^{(1)}$, $A^{(2)}$, $\cdots$, $A^{(n-2)}$, $A^{(n-1,1)}$ and $A^{(n,1)}$.
Given these
measurements and the initial state, we can calculate the Mermin value $\mathbf{I}^{(k)}_{n}(\theta)$  with respect to the state $\rho^{(k)}$ and the observables $\{M_{x_i}^{(i)}\}_{x_i\in\mathbb{Z}_2}$ for $i=1,2,\cdots, n-2$  and  $\{M_{x_i}^{(i,k)}\}_{x_i\in\mathbb{Z}_2}$ for $i=n-1,n$   as follows (see Appendix \ref{ap6} for the detailed calculation) 
$$ \mathbf{I}^{(k)}_{n}(\theta) =\displaystyle N_n\frac{\theta\gamma_k(\theta)+\theta  P_k(\theta)}{2^{k-1}}, $$
where \begin{equation}\label{eq:Constant2}
    N_n=\begin{cases}
        (\sqrt{2})^{n-4}(-1)^ {\lfloor \frac{n}{4}\rfloor+1},  &\text{if } n \equiv 0   \mod 4,\\[2mm](\sqrt{2})^{n-5}(-1)^ {\lfloor \frac{n}{4}\rfloor},  &\text{if } n \equiv 1   \mod 4,\\[2mm](\sqrt{2})^{n-4}(-1)^ {\lfloor \frac{n}{4}\rfloor},  &\text{if } n \equiv 2   \mod 4,\\[2mm](\sqrt{2})^{n-5}(-1)^ {\lfloor \frac{n}{4}\rfloor+1},  &\text{if } n \equiv 3   \mod 4.
    \end{cases}
\end{equation}

From Eq.  \eqref{eq:Constant2}, it's easy to check that $|N_n|\geq1$  whenever $n\geq 4$. Then, using Lemma \ref{Lemma},  we can deduce the following theorem.

\begin{theorem}\label{thm:Main3}
    For $\forall$ $K\in\mathbb{N}$, there exist some $\theta\in(-1,1)$ and a sequence $\{\gamma_{k}(\theta)\}_{k=1}^{n}$  such that $\gamma_k(\theta)\in(0,1)$ and  $\mathbf{I}^{(k)}_{n}(\theta) >1$  for $k\in\{1,2,\cdots,K\}$. That is,  we can detect unbounded sequential 
$n$-partite standard nonlocality  starting from  $\mathrm{GHZ}$ state in the double-chain setting.
\end{theorem}

\section{CONCLUSIONS AND DISCUSSION}\label{s8}
We have designed appropriate measurements that facilitate the detection of unbounded sequential multipartite nonlocality through the violation of the Mermin inequality. The choice of measurement is crucial as it can enhance the degree of violation. By refining the measurements introduced in \cite{19}, we have shown that an arbitrary number of independent Charlies can observe standard tripartite nonlocality with a single Alice and a single Bob for an initially shared W state. Following this, we inferred the existence of unbounded sequential $n$-partite nonlocality for the initially shared GHZ state in the single-chain scenario, as demonstrated by the Mermin inequality. Furthermore, we introduce the new double-chain scenario, and we reached the same conclusion in the $n$-partite scenario.

Multipartite quantum nonlocality is a vital resource for both foundational quantum mechanics research and quantum communication applications. In \cite{25}, the authors experimentally illustrate that the multipartite nonlocality is more ubiquitous than people have ever thought and would be useful for studying multipartite nonlocality in the scenarios of practically realizing multipartite quantum communication tasks among distant parties. Additionally, nonlocality can be explored in various other scenarios, such as \cite{26}, star network scenarios \cite{27,28} and in a quantum network \cite{29}, among others.

It is important to note that several issues remain open for further investigation. In the single-chain and double-chain scenario, it is currently unknown whether a concrete measurement can be identified that would allow an arbitrary number of independent observers to detect sequential nonlocality in $n$-partite systems via the Mermin inequality for an initially shared W state. Another consideration is whether in an arbitrary chain setting, we can also be used to detect unbounded sequential $n$-partite nonlocality. These questions merit further exploration to deepen our understanding of multipartite nonlocality and its potential applications.
\bigskip
\begin{center}
    {\noindent{\bf ACKNOWLEDGMENTS}\, \,}
\end{center}

This work was supported by National Natural Science Foundation of China  under Grant No. 12371458, the Guangdong Basic and Applied Basic Research Foundation under Grants No. 2023A1515012074   and No. 2024A1515010380.

\bigskip

\onecolumngrid

\appendix
\section{THE CALCULATION OF THE COEFFICIENTS OF THE MERMIN POLYNOMIALS $c_{\mathbf{v}}$ IN $n$-PARTITE SYSTEMS}\label{ap1}
Using Eq.(\ref{m5}), we know that
\begin{equation}\label{k1}
     \begin{array}{c}
 \displaystyle  M_n=\sum_{\mathbf{v}\in \mathbb{Z}_2^n} c_{\mathbf{v}} \prod_{i=1}^n M_{v_i}^{(i)}=\sum_{\mathbf{v}\in \mathbb{Z}_2^{n-1}} c_{(\mathbf{v},0)} \prod_{i=1}^{n-1} M_{v_i}^{(i)}M_{0}^{(n)}+\sum_{\mathbf{v}\in \mathbb{Z}_2^{n-1}} c_{(\mathbf{v},1)} \prod_{i=1}^{n-1} M_{v_i}^{(i)}M_{1}^{(n)},\\[2mm] 
    \displaystyle  M'_n=\sum_{\mathbf{v}\in \mathbb{Z}_2^n} c'_{\mathbf{v}} \prod_{i=1}^n M_{v_i}^{(i)}=\sum_{\mathbf{v}\in \mathbb{Z}_2^{n-1}} c'_{(\mathbf{v},0)} \prod_{i=1}^{n-1} M_{v_i}^{(i)}M_{0}^{(n)}+\sum_{\mathbf{v}\in \mathbb{Z}_2^{n-1}} c'_{(\mathbf{v},1)} \prod_{i=1}^{n-1} M_{v_i}^{(i)}M_{1}^{(n)}.
      \end{array}
\end{equation}
Using Eq.(\ref{m1}), we have
\begin{equation}\label{k3}
  \begin{array}{l}
 \displaystyle  M_n=\frac{M_{n-1}+M'_{n-1}}{2}M_{0}^{(n)}+\frac{M_{n-1}-M'_{n-1}}{2}M_{1}^{(n)},\ \

 \displaystyle  M'_n=\frac{M'_{n-1}-M_{n-1}}{2}M_{0}^{(n)}+\frac{M_{n-1}+M'_{n-1}}{2}M_{1}^{(n)},
   \end{array}
\end{equation}
 with $M_1=M_{0}^{(1)}$, $M'_1=M_{1}^{(1)}$.

Let $\mathbf{v}=(v_1,v_2,\cdots,v_{n-1})\in{\mathbb{Z}_2^{n-1}}$, $v_n\in{\mathbb{Z}_2}$, through the four equations above, we can deduce that
\begin{equation}
    c_{(\mathbf{v},0)}=\frac{c_{\mathbf{v}}+c'_{\mathbf{v}}}{2}, c_{(\mathbf{v},1)}=\frac{c_{\mathbf{v}}-c'_{\mathbf{v}}}{2}, c'_{(\mathbf{v},0)}=\frac{-c_{\mathbf{v}}+c'_{\mathbf{v}}}{2}, c'_{(\mathbf{v},1)}=\frac{c_{\mathbf{v}}+c'_{\mathbf{v}}}{2}.
\end{equation}
Then, we have
\begin{equation}\label{j1}
\begin{pmatrix}
    c_{(\mathbf{v},0)}\\c'_{(\mathbf{v},0)}
\end{pmatrix}=\begin{pmatrix}
\frac{1}{2} &\frac{1}{2}\\-\frac{1}{2}& \frac{1}{2}
\end{pmatrix}
\begin{pmatrix}
    c_{\mathbf{v}}\\c'_{\mathbf{v}}
\end{pmatrix},\begin{pmatrix}
    c_{(\mathbf{v},1)}\\c'_{(\mathbf{v},1)}
\end{pmatrix}=\begin{pmatrix}
\frac{1}{2} &-\frac{1}{2}\\\frac{1}{2}& \frac{1}{2}
\end{pmatrix}
\begin{pmatrix}
    c_{\mathbf{v}}\\c'_{\mathbf{v}}
\end{pmatrix}.
\end{equation}
Denote $
    H=\frac{1}{\sqrt{2}}\begin{pmatrix}
1 &1\\-1& 1
\end{pmatrix},$
we can get \begin{equation}\label{}
\begin{pmatrix}
    c_{(\mathbf{v},0)}\\c'_{(\mathbf{v},0)}
\end{pmatrix}=\frac{1}{\sqrt{2}}H^{(-1)^{0}}
\begin{pmatrix}
    c_{\mathbf{v}}\\c'_{\mathbf{v}}
\end{pmatrix},\begin{pmatrix}
    c_{(\mathbf{v},1)}\\c'_{(\mathbf{v},1)}
\end{pmatrix}=\frac{1}{\sqrt{2}}H^{(-1)^{1}}
\begin{pmatrix}
    c_{\mathbf{v}}\\c'_{\mathbf{v}}
\end{pmatrix}.
\end{equation}
Hence, \begin{equation}\label{}
\begin{pmatrix}
    c_{(\mathbf{v},v_n)}\\c'_{(\mathbf{v},v_n)}
\end{pmatrix}=\frac{1}{\sqrt{2}}H^{(-1)^{v_n}}
\begin{pmatrix}
    c_{\mathbf{v}}\\c'_{\mathbf{v}}
\end{pmatrix}=\cdots=(\frac{1}{\sqrt{2}})^{n-1}H^{(-1)^{v_n}+\cdots+(-1)^{v_2}}\begin{pmatrix}
    c_{\mathbf{v_1}}\\c'_{\mathbf{v_1}}
\end{pmatrix}.
\end{equation}
Since \begin{equation}\label{}
\begin{pmatrix}
    c_{0}\\c'_{0}
\end{pmatrix}=\begin{pmatrix}
1 \\0
\end{pmatrix}
=H^{(-1)^0}\begin{pmatrix}
\frac{1}{\sqrt{2}} \\\frac{1}{\sqrt{2}}
\end{pmatrix}, \begin{pmatrix}
    c_{1}\\c'_{1}
\end{pmatrix}=\begin{pmatrix}
0 \\1
\end{pmatrix}=H^{(-1)^1}\begin{pmatrix}
\frac{1}{\sqrt{2}} \\\frac{1}{\sqrt{2}}
\end{pmatrix},
\end{equation}
then \begin{equation}\label{j1}
\begin{pmatrix}
    c_{(\mathbf{v},v_n)}\\c'_{(\mathbf{v},v_n)}
\end{pmatrix}=\cdots=(\frac{1}{\sqrt{2}})^{n-1}H^{(-1)^{v_n}+\cdots+(-1)^{v_2}+(-1)^{v_1}}\begin{pmatrix}
\frac{1}{\sqrt{2}} \\\frac{1}{\sqrt{2}}
\end{pmatrix}=(\frac{1}{\sqrt{2}})^{n-1}H^{n-2|(\mathbf{v},v_n)|}\begin{pmatrix}
\frac{1}{\sqrt{2}} \\\frac{1}{\sqrt{2}}
\end{pmatrix}.
\end{equation}

That is, for any $\mathbf{v}=(v_1,v_2,\cdots,v_n)\in \mathbb{Z}_2^n$, we have 
\begin{equation}\label{j1}
\begin{pmatrix}
    c_{\mathbf{v}}\\c'_{\mathbf{v}}
\end{pmatrix}=(\frac{1}{\sqrt{2}})^{n-1}H^{n-2|\mathbf{v}|}\begin{pmatrix}
\frac{1}{\sqrt{2}} \\\frac{1}{\sqrt{2}}
\end{pmatrix}.\end{equation}
We can find out the eigenvalues and eigenvector of $H$. In fact, 
 $$ H\begin{pmatrix}
\frac{-i}{\sqrt{2}} \\\frac{1}{\sqrt{2}}
\end{pmatrix}= e^{\pi  i/4}\begin{pmatrix}
\frac{-i}{\sqrt{2}} \\\frac{1}{\sqrt{2}}
\end{pmatrix},
\ \ \ \ \text{and} \ \ \ 
H\begin{pmatrix}
\frac{i}{\sqrt{2}} \\\frac{1}{\sqrt{2}}
\end{pmatrix}= e^{-\pi  i/4}\begin{pmatrix}
\frac{i}{\sqrt{2}} \\\frac{1}{\sqrt{2}}
\end{pmatrix}.$$
Let $\lambda_1=e^{\pi  i/4}$ and  $\lambda_2=e^{-\pi  i/4}.$ As we have the decomposition
$$\begin{pmatrix}
\frac{1}{\sqrt{2}} \\\frac{1}{\sqrt{2}}
\end{pmatrix}= \frac{1-i}{2} \begin{pmatrix}
\frac{i}{\sqrt{2}} \\\frac{1}{\sqrt{2}}
\end{pmatrix}+\frac{1+i}{2} \begin{pmatrix}
\frac{-i}{\sqrt{2}} \\\frac{1}{\sqrt{2}}
\end{pmatrix}, $$
so we obtain that
\begin{equation}\label{j1}
\begin{pmatrix}
    c_{\mathbf{v}}\\c'_{\mathbf{v}}
\end{pmatrix}=\begin{pmatrix}
(\frac{1}{\sqrt{2}})^{n-1} \lambda_2^{n-2|\mathbf{v}|} \frac{1+i}{2\sqrt{2}} + (\frac{1}{\sqrt{2}})^{n-1} \lambda_1^{n-2|\mathbf{v}|} \frac{1-i}{2\sqrt{2}} \\[3mm]
(\frac{1}{\sqrt{2}})^{n-1} \lambda_2^{n-2|\mathbf{v}|} \frac{1-i}{2\sqrt{2}} + (\frac{1}{\sqrt{2}})^{n-1} \lambda_1^{n-2|\mathbf{v}|} \frac{1+
i}{2\sqrt{2}}
\end{pmatrix}.\end{equation}
Therefore,  as $\lambda_1^{-1}=\lambda_2 =\frac{1-i}{\sqrt{2}},$ we have 
\begin{equation}\label{eq:c_Expression}
    c_{\mathbf{v}}=(\frac{1}{\sqrt{2}})^{n-1} \lambda_2^{n-2|\mathbf{v}|} \frac{1+i}{2\sqrt{2}} + (\frac{1}{\sqrt{2}})^{n-1} \lambda_1^{n-2|\mathbf{v}|} \frac{1-i}{2\sqrt{2}}=\frac{1}{2} \left((\frac{1}{\sqrt{2}}\lambda_2)^{n-1} \lambda_1^{2|\mathbf{v}|}+(\frac{1}{\sqrt{2}}\lambda_1)^{n-1} \lambda_2^{2|\mathbf{v}|}\right).
\end{equation}  

\section{THE CALCULATION OF $\mathbf{I}^{(k)}_{3}(\theta)$ AMONG ALICE BOB AND CHARLIE$^{(k)}$}\label{ap2}
In this measurement strategy, Alice's observables are defined by
\begin{eqnarray*}\label{eq:A1}
  M^{(1)}_0=\sin(\theta)\sigma_{1}+\cos(\theta)\sigma_{3}, \ \
  M^{(1)}_1=   \sin(\theta)\sigma_{1} -\cos(\theta)\sigma_{3}.
\end{eqnarray*}
Bob's observables are defined by
\begin{eqnarray*}\label{eq:A1}
  M^{(2)}_0=\sin(\theta)\sigma_{1}+\cos(\theta)\sigma_{3}, \ \
  M^{(2)}_1=   \sin(\theta)\sigma_{1} -\cos(\theta)\sigma_{3}.
\end{eqnarray*}
 Charlie$^{(k)}$'s $(1 \leq k \leq K)$  observables are defined by
\begin{eqnarray*}\label{eq:B1}
  M^{(3,k)}_0=\sigma_{3}, \ \
  M^{(3,k)}_1= \gamma_k(\theta)\sigma_{1},  \ \  (0<\gamma_k(\theta)<1).
\end{eqnarray*}
Let $\rho^{(k-1)}$ shared by Alice, Bob and Charlie$^{(k-1)}$ prior to Charlie$^{(k-1)}$'s measurements, and with an initial state: $\rho^{(1)}=\left|\psi\right\rangle\left\langle\psi\right|$, 
$\left|\psi\right\rangle=\frac{1}{\sqrt{3}}(\left|100\right\rangle+\left|010\right\rangle+\left|001\right\rangle)$. Using the L\"{u}ders rule, the state sent to Charlie$^{(k)}$ is
$$\begin{array}{rl}
   \rho{^{(k)}} & = \frac{1}{2}\sum\limits_{c,z}(\mathbb{I}\otimes\mathbb{I}\otimes\sqrt{M^{(3,k-1)}_{c|z}}\rho^{(k-1)}\mathbb{I}\otimes\mathbb{I}\otimes\sqrt{M^{(3,k-1)}_{c|z}})\\[3mm]
   & =\frac{1}{2}(\mathbb{I}\otimes\mathbb{I}\otimes\frac{\mathbb{I}+\sigma_{3}}{2})\rho^{(k-1)}(\mathbb{I}\otimes\mathbb{I}\otimes\frac{\mathbb{I}+\sigma_{3}}{2})+(\mathbb{I}\otimes\mathbb{I}\otimes\frac{\mathbb{I}-\sigma_{3}}{2})\rho^{(k-1)}(\mathbb{I}\otimes\mathbb{I}\otimes\frac{\mathbb{I}-\sigma_{3}}{2})\\[3mm]
&+(\mathbb{I}\otimes\mathbb{I}\otimes\sqrt{\frac{\mathbb{I}+\gamma_{k-1}(\theta)\sigma_{1}}{2}})\rho^{(k-1)}(\mathbb{I}\otimes\mathbb{I}\otimes\sqrt{\frac{\mathbb{I}+\gamma_{k-1}(\theta)\sigma_{1}}{2}})+ (\mathbb{I}\otimes\mathbb{I}\otimes\sqrt{\frac{\mathbb{I}-\gamma_{k-1}(\theta)\sigma_{1}}{2}})\rho^{(k-1)}(\mathbb{I}\otimes\mathbb{I}\otimes\sqrt{\frac{\mathbb{I}-\gamma_{k-1}(\theta)\sigma_{1}}{2}}) \\[3mm]
& =\frac{2+\sqrt{1-\gamma_{k-1}^{2}(\theta)}}{4}\rho^{(k-1)}+\frac{1}{4}(\mathbb{I}\otimes\mathbb{I}\otimes\sigma_{3})\rho^{(k-1)}(\mathbb{I}\otimes\mathbb{I}\otimes\sigma_{3})\\[3mm]
&+\frac{1-\sqrt{1-\gamma_{k-1}^{2}(\theta)}}{4}(\mathbb{I}\otimes\mathbb{I}\otimes\sigma_{1})\rho^{(k-1)}(\mathbb{I}\otimes\mathbb{I}\otimes\sigma_{1}),
\end{array}$$
where we use the identity for the final calculation
\begin{equation}\label{g1}
\sqrt{\frac{\mathbb{I}\pm\gamma_{k}(\theta)\sigma_{\vec{r}}}{2}}=\frac{(\sqrt{1+\gamma_{k}(\theta)}+\sqrt{1-\gamma_{k}})\mathbb{I}\pm(\sqrt{1+\gamma_{k}(\theta)}-\sqrt{1-\gamma_{k}(\theta)})\sigma_{\vec{r}}}{2\sqrt{2}}.
\end{equation}
As 
\begin{equation}
\begin{array}{l}
    \displaystyle\mathrm{Tr}[\rho^{(k)}(M_1^{(1)}M_0^{(2)}M_0^{(3,k)})]=\mathrm{Tr}[\rho^{(k)}(-\cos{\theta}\sigma_{3}+\sin{\theta}\sigma_{1})\otimes (\cos{\theta}\sigma_{3}+\sin{\theta}\sigma_{1})\otimes\sigma_{3}]=\frac{P_{k}(\theta)}{2^{k-1}}(\cos^2{\theta}+\frac{2\sin^2{\theta}}{3}),\\[2mm] 
    \displaystyle \mathrm{Tr}[\rho^{(k)}(M_0^{(1)}M_1^{(2)}M_0^{(3,k)})]=\mathrm{Tr}[\rho^{(k)}(\cos{\theta}\sigma_{3}+\sin{\theta}\sigma_{1})\otimes (-\cos{\theta}\sigma_{3}+\sin{\theta}\sigma_{1})\otimes\sigma_{3}]=\frac{P_{k}(\theta)}{2^{k-1}}(\cos^2{\theta}+\frac{2\sin^2{\theta}}{3}),\\[2mm] 
    \displaystyle\mathrm{Tr}[\rho^{(k)}(M_0^{(1)}M_0^{(2)}M_1^{(3,k)})]=\mathrm{Tr}[\rho^{(k)}(\cos{\theta}\sigma_{3}+\sin{\theta}\sigma_{1})\otimes (\cos{\theta}\sigma_{3}+\sin{\theta}\sigma_{1})\otimes(\gamma_{k}(\theta)\sigma_{1})]=\frac{\gamma_{k}(\theta)}{2^{k-1}}(\frac{4\sin{\theta}\cos{\theta}}{3}),\\[2mm] 
    \displaystyle-\mathrm{Tr}[\rho^{(k)}(M_1^{(1)}M_1^{(2)}M_1^{(3,k)})]=\mathrm{Tr}[\rho^{(k)}(-\cos{\theta}\sigma_{3}+\sin{\theta}\sigma_{1})\otimes (-\cos{\theta}\sigma_{3}+\sin{\theta}\sigma_{1})\otimes(\gamma_{k}(\theta)\sigma_{1})]=\frac{\gamma_{k}(\theta)}{2^{k-1}}(\frac{4\sin{\theta}\cos{\theta}}{3}),
    \end{array}
\end{equation}
 we  could calculate  the Mermin value $\mathbf{I}^{(k)}_{3}(\theta)$ of $\rho^{(k)}$:
\begin{equation}
    \mathbf{I}^{(k)}_{3}(\theta)
=\frac{P_{k}(\theta)}{2^{k}}(2\cos^2{\theta}+\frac{4\sin^2{\theta}}{3})+\frac{\gamma_{k}(\theta)}{2^{k}}(\frac{8\sin{\theta}\cos{\theta}}{3}),
\end{equation}  
where $P_{k}(\theta)=\prod\limits_{j=1}^{k-1}(1+\sqrt{1-\gamma_{j}^{2}(\theta)})$.
In particular, $\mathbf{I}^{(1)}_{3}(\theta)=(\cos^2{\theta}+\frac{2\sin^2{\theta}}{3})+(\frac{4\sin{\theta}\cos{\theta}{\gamma_{k}}}{3})$.
\section{THE PROOF OF THEOREM 1}\label{ap3}
For the given measurements in Sec. \ref{s3}, in order to $\mathbf{I}^{(k)}_{3}(\theta)>1$ with $1 \leq k \leq K$, we have to find out right $\gamma_k$ and $\theta$ such that 
\begin{eqnarray}\label{}
\gamma_k(\theta)>\frac{2^k-P_k(\theta)(2-\frac{2\sin^2{(\theta)}}{3})}{\frac{8\sin(\theta)\cos(\theta)}{3}}.
\end{eqnarray}
To achieve it, for $\forall\epsilon>0$, we define a specfic sequence $\{\gamma_k(\theta)\}_{k=1}^K$
\begin{equation}\label{}
  \gamma_{k}(\theta)=\left\{
                       \begin{array}{ll}
                         (1+\epsilon)(\frac{\tan(\theta)}{4}), & \hbox{$k=1$;} \\
                     (1+\epsilon)(\frac{(2^k-P_k(\theta))(2-\frac{2\sin^2{(\theta)}}{3})}{\frac{8\sin(\theta)\cos(\theta)}{3}}) , & \hbox{$0\leq\gamma_{k-1}(\theta)\leq1$ ;} \\
                         \infty, & \hbox{otherwise.}
                       \end{array}
                     \right.
\end{equation}
Then we can get
\begin{equation}
  \frac{\gamma_{k}(\theta)}{\gamma_{k-1}(\theta)}>2\Leftrightarrow \frac{P_{k-1}(\theta)}{P_k(\theta)}>\frac{1}{2}\Leftrightarrow 0<\gamma_{k-1}(\theta)\leq1.
\end{equation}
Here $\gamma_{1}(\theta)=(1+\epsilon)(\frac{\tan(\theta)}{4})$, and $\lim\limits_{\theta\rightarrow 0^{+}}\gamma_{1}(\theta)=0.$
 By the induction, we can suppose there exists a $\theta_{k-1}$ such that on the interval $ (0,\theta_{k-1})$, all $\gamma_{i}(\theta)\in(0,1),$  and $\lim\limits_{\theta\rightarrow 0^{+}}\gamma_{i}(\theta)=0$  for $i=1,2,\cdots ,k-1.$ Meanwhile, we note that $P_k(\theta)$ as a function on the small interval $(0,\theta_{k-1})$, the limit of its  differetial at 0$^+$ could be calculated as
$$
 \lim\limits_{\theta\rightarrow 0^{+}}P'_k(\theta)=\lim\limits_{\theta\rightarrow 0^{+}}\sum_{j=1}^{k-1}\left( \frac{-2 \gamma_j(\theta) \gamma_j'(\theta)}{2\sqrt{1-\gamma_j^2(\theta)}}\right) \frac{P_k(\theta)}{1+\sqrt{1-\gamma_j^2(\theta)}}=0.
$$
 Then according to the definition of $\gamma_{k}(\theta)$, we will have $$\lim\limits_{\theta\rightarrow 0^{+}}\gamma_{k}(\theta)=\lim\limits_{\theta\rightarrow 0^{+}} (1+\epsilon)(\frac{(2^k-P_k(\theta))(2-\frac{2\sin^2{(\theta)}}{3})}{\frac{8\sin(\theta)\cos(\theta)}{3}}) =\lim\limits_{\theta\rightarrow 0^{+}}\frac{-2P'_k(\theta)+\frac{2\sin{2\theta}P_k(\theta)}{3}+\frac{2\sin^2{\theta}P'_k(\theta)}{3}}{\frac{8\cos{2\theta}}{3}}=0,$$
here we use the limit $\lim\limits_{\theta\rightarrow 0^{+}} P_k(\theta)=2^{k-1} $ and L'Hopspital rule.   So  $\forall K\in\mathbb{N},$ we can find a $\theta_{K}\in(0,1)$ such that $0<\gamma_{1}(\theta)<\gamma_{2}(\theta)<\cdots<\gamma_{K}(\theta)<1$ for all $\theta\in(0,\theta_{K})$.
\section{THE CALCULATION OF $\mathbf{I}^{(k)}_{n}(\theta)$ IN SINGLE-CHAIN SCENARIO}\label{ap4}
In this measurement strategy,  $A^{(j)}$'s observables are defined by
\begin{eqnarray}\label{eq:A1}
 M^{(j)}_0 =\sigma_{1}, \ \
  M^{(j)}_1 = \sigma_{2}.
\end{eqnarray}
where $j=1,2,\cdots,n-2$.
 $A^{(n-1)}$'s observables are defined by
\begin{eqnarray}\label{eq:A1}
 M^{(n-1)}_0 =\theta\sigma_{1}, \ \
  M^{(n-1)}_1 = \theta\sigma_{2}.
\end{eqnarray}
$A^{(n,k)}$'s $(1\leq k\leq K)$ observables are defined by
\begin{eqnarray}\label{eq:A1}
 M^{(n,k)}_0 =\sigma_{1}, \ \
  M^{(n,k)}_1 =  {\gamma}_{k}(\theta)\sigma_{2},\ \  (0< {\gamma}_{k}(\theta)<1).
\end{eqnarray}  

 Let $\rho^{(k-1)}$ shared by $A^{(1)}$,  $A^{(2)}$, $\cdots$,  $A^{(n)}$, and $A^{(n,k-1)}$ prior to $A^{(n,k-1)}$'s measurements and with an initial state:  $\rho^{(1)}=\left|\mathrm{GHZ}_n\right\rangle\left\langle\mathrm{GHZ}_n\right|$, 
$\left|\mathrm{GHZ}_n\right\rangle=\frac{1}{\sqrt{2}}(\left|00\cdots0\right\rangle+\left|11\cdots1\right\rangle)$. Using the L\"{u}ders rule, the state sent to $A^{(n,k)}$ is
 \begin{equation}\label{eq:recursive}
\begin{array}{rl}
   \rho^{(k)} & =\frac{1}{2}\sum\limits(\mathbb{I}^{\otimes (n-1)}\otimes\sqrt{M^{(n,k-1)}_{a^{(n,k-1)}|x^{(n,k-1)}}})\rho^{(k-1)}(\mathbb{I}^{\otimes (n-1)}\otimes\sqrt{M^{(n,k-1)}_{a^{(n,k-1)}|x^{(n,k-1)}}}),\\[3mm]
   & =\frac{1}{2}(\mathbb{I} ^{\otimes (n-1)}\otimes\frac{\mathbb{I}+\sigma_{1}}{2})\rho^{(k-1)}(\mathbb{I} ^{\otimes (n-1)}\otimes\frac{\mathbb{I}+\sigma_{1}}{2})+(\mathbb{I} ^{\otimes (n-1)}\otimes\frac{\mathbb{I}-\sigma_{1}}{2})\rho^{(k-1)}(\mathbb{I} ^{\otimes (n-1)}\otimes\frac{\mathbb{I}-\sigma_{1}}{2})\\[3mm]
&+(\mathbb{I} ^{\otimes (n-1)}\otimes\sqrt{\frac{\mathbb{I}+ {\gamma}_{k-1}(\theta)\sigma_{2}}{2}})\rho^{(k-1)}(\mathbb{I} ^{\otimes (n-1)}\otimes\sqrt{\frac{\mathbb{I}+ {\gamma}_{k-1}(\theta)\sigma_{2}}{2}})+ (\mathbb{I} ^{\otimes (n-1)}\otimes\sqrt{\frac{\mathbb{I}- {\gamma}_{k-1}(\theta)\sigma_{2}}{2}})\rho^{(k-1)}(\mathbb{I} ^{\otimes (n-1)}\otimes\sqrt{\frac{\mathbb{I}- {\gamma}_{k-1}(\theta)\sigma_{2}}{2}}) \\[3mm]
& =\frac{2+\sqrt{1- {\gamma}_{k-1}^{2}(\theta)}}{4}\rho^{(k-1)}+\frac{1}{4}(\mathbb{I} ^{\otimes (n-1)}\otimes\sigma_{1})\rho^{(k-1)}(\mathbb{I} ^{\otimes (n-1)}\otimes\sigma_{1})+\frac{1-\sqrt{1- {\gamma}_{k-1}^{2}(\theta)}}{4}(\mathbb{I} ^{\otimes (n-1)}\otimes\sigma_{2})\rho^{(k-1)}(\mathbb{I} ^{\otimes (n-1)}\otimes\sigma_{2}),
\end{array}    
\end{equation}
where we use Eq. (\ref{g1}).

To calculate $\mathbf{I}^{(k)}_{n}(\theta)$, by definition
\begin{equation}\label{eq:Mermin_P}
    \mathbf{I}^{(k)}_{n}(\theta)=\sum_{\mathbf{v}\in \mathbb{Z}_2^n} c_\mathbf{v} \mathrm{Tr}[\rho^{(k)}(M_{v_1}^{(1)}M_{v_2}^{(2)}\cdots M_{v_n}^{(n,k)})],
\end{equation} 
we should   calculate out each term $\mathrm{Tr}[\rho^{(k)}(M_{v_1}^{(1)}M_{v_2}^{(2)}\cdots M_{v_n}^{(n,k)})].$ For any vector  $\mathbf{v}=(v_1,v_2,\cdots,v_{n})\in \mathbb{Z}_2^{n},$   we define 
$\Tilde{\mathbf{v}}=(\Tilde{v}_1,\Tilde{v}_2,\cdots,\Tilde{v}_{n}):=(v_1+1,v_2+1,\cdots, v_n+1)\in \{1,2\}^n.$
Note that 
$$ \left(\sigma_{\Tilde{v}_1} \otimes \sigma_{\Tilde{v}_2}\otimes \cdots \otimes \sigma_{\Tilde{v}_n}\right) |\mathrm{GHZ}_n\rangle =\frac{1}{\sqrt{2}}( i^{|\mathbf{v}|} |11\cdots 1\rangle+(-i)^{|\mathbf{v}|}|00\cdots 0\rangle ).$$
Therefore, 
$$\mathrm{Tr}[\rho^{(1) } \left(\sigma_{\Tilde{v}_1} \otimes \sigma_{\Tilde{v}_2}\otimes \cdots \otimes \sigma_{\Tilde{v}_n}\right) ]= \frac{(-i)^{|\mathbf{v}|}+ i^{|\mathbf{v}|}}{2}=\begin{cases}
    1,  & \text{if } \mathbf{v}\equiv 0 \mod 4\\
    -1,  &\text{if } \mathbf{v}\equiv 2 \mod 4\\
    0,   & \text{otherwise. }
\end{cases}$$ 

For each $\mathbf{v}=(v_1,v_2,\cdots,v_{n})\in \mathbb{Z}_2^{n},$ by substituting the $\rho^{(k)}$ with Eq. \eqref{eq:recursive}, one finds that
\begin{equation}\label{eq:recursive3cases}
    \mathrm{Tr}[\rho^{(k)} \left(\sigma_{\Tilde{v}_1} \otimes \sigma_{\Tilde{v}_2}\otimes \cdots \otimes \sigma_{\Tilde{v}_n}\right)]=\begin{cases}
    \frac{1+\sqrt{1-\gamma_{k-1}^2(\theta)}}{2}\mathrm{Tr}[\rho^{(k-1)} \left(\sigma_{\Tilde{v}_1} \otimes \sigma_{\Tilde{v}_2}\otimes \cdots \otimes \sigma_{\Tilde{v}_n}\right)],  &\text{if } v_{n}=0,\\[2mm]
    \frac{1}{2}
    \mathrm{Tr}[\rho^{(k-1)} \left(\sigma_{\Tilde{v}_1} \otimes \sigma_{\Tilde{v}_2}\otimes \cdots \otimes \sigma_{\Tilde{v}_n}\right)],  & \text{if } v_{n}=1.
   \end{cases}
\end{equation} 
Repeating this recursive equation $(k-1)$ times,  one has that  
\begin{equation}\label{eq:initial3cases}
  \mathrm{Tr}[\rho^{(k)} \left(\sigma_{\Tilde{v}_1} \otimes \sigma_{\Tilde{v}_2}\otimes \cdots \otimes \sigma_{\Tilde{v}_n}\right)]= \begin{cases} 
  \frac{P_k(\theta)}{2^{k-1}}\frac{(-i)^{|\mathbf{v}|}+ i^{|\mathbf{v}|}}{2},  &\text{if } v_{n}=0,\\[2mm]
  \frac{1}{2^{k-1}}
   \frac{(-i)^{|\mathbf{v}|}+ i^{|\mathbf{v}|}}{2},  &\text{if } v_{n}=1.
\end{cases}
\end{equation}
One finds that 
\begin{equation}\label{eq:Maininitial3cases} 
    \mathrm{Tr}[\rho^{(k)}(M^{(1)}_{v_1}M^{(2)}_{v_2}\cdots M^{(n,k)}_{v_{n} })]=\begin{cases} \theta  \mathrm{Tr}[\rho^{(k)}\left(\sigma_{\Tilde{v}_1} \otimes \sigma_{\Tilde{v}_2}\otimes \cdots \otimes \sigma_{\Tilde{v}_n}\right)],  &\text{if } v_{n}=0,\\[2mm]
   \theta  \gamma_k(\theta) \mathrm{Tr}[\rho^{(k)}\left(\sigma_{\Tilde{v}_1} \otimes \sigma_{\Tilde{v}_2}\otimes \cdots \otimes \sigma_{\Tilde{v}_n}\right)],  &\text{if } v_{n}=1.
\end{cases}
\end{equation}
Therefore,  by Eqs. \eqref{eq:Mermin_P}, \eqref{eq:initial3cases}, and \eqref{eq:Maininitial3cases}, the Mermin value $\mathbf{I}^{(k)}_{n}(\theta)$ of $\rho^{(k)}$ can be written as  
$$ \mathbf{I}^{(k)}_{n}(\theta) =\left(\sum_{\mathbf{v}\in \mathbb{Z}_2^n, v_n=0} c_\mathbf{v} \frac{(-i)^{|\mathbf{v}|}+ i^{|\mathbf{v}|}}{2}  \right)\frac{\theta P_k(\theta)}{2^{k-1}}+\left(\sum_{\mathbf{v}\in \mathbb{Z}_2^n, v_n=1} c_\mathbf{v} \frac{(-i)^{|\mathbf{v}|}+ i^{|\mathbf{v}|}}{2}  \right)\frac{\theta  \gamma_k(\theta)}{2^{k-1}}. $$
Now we calculate the two coefficients in the above equation ($\lambda_1^2=i$ and $\lambda_2^2=-i $)
\begin{equation}\label{eq:Coeff3case1}
\begin{array}{rcl}
   &&\displaystyle\sum_{\mathbf{v}\in \mathbb{Z}_2^n, v_n=0} c_\mathbf{v} \frac{(-i)^{|\mathbf{v}|}+ i^{|\mathbf{v}|}}{2}\\[4mm] &=&\displaystyle\frac{1}{4}\sum_{\mathbf{v}\in \mathbb{Z}_2^n, v_n=0}  \left((\frac{1}{\sqrt{2}}\lambda_2)^{n-1} \lambda_1^{2|\mathbf{v}|}+(\frac{1}{\sqrt{2}}\lambda_1)^{n-1} \lambda_2^{2|\mathbf{v}|}\right)\left((-i)^{|\mathbf{v}|}+ i^{|\mathbf{v}|}\right)\\[4mm]
   &=& \displaystyle \frac{1}{4}\sum_{\mathbf{v}\in \mathbb{Z}_2^n, v_n=0, |\mathbf{v}|=k}    \left((\frac{1}{\sqrt{2}}\lambda_2)^{n-1} \lambda_1^{2k}+(\frac{1}{\sqrt{2}}\lambda_1)^{n-1} \lambda_2^{2k}\right)\left((-i)^{k}+ i^{k}\right)\\[4mm]
   &=& \displaystyle \frac{1}{4}\sum_{k=0}^{n-1}  \binom{n-1}{k}  \left((\frac{1}{\sqrt{2}}\lambda_2)^{n-1} \lambda_1^{2k}+(\frac{1}{\sqrt{2}}\lambda_1)^{n-1} \lambda_2^{2k}\right)\left((-i)^{k}+ i^{k}\right)\\[4mm]
   &=& \displaystyle \frac{1}{4}\sum_{k=0}^{n-1}  \binom{n-1}{k}  \left((\frac{1}{\sqrt{2}}\lambda_2)^{n-1} \lambda_1^{2k}+(\frac{1}{\sqrt{2}}\lambda_1)^{n-1} \lambda_2^{2k}\right)\left((-i)^{k}+ i^{k}\right)\\[4mm]
    &=& \displaystyle \frac{(\frac{1}{\sqrt{2}}\lambda_2)^{n-1}}{4}\sum_{k=0}^{n-1}  \binom{n-1}{k}    \lambda_1^{2k} \left((-i)^{k}+ i^{k}\right)+ \displaystyle \frac{(\frac{1}{\sqrt{2}}\lambda_1)^{n-1}}{4}\sum_{k=0}^{n-1}  \binom{n-1}{k}    \lambda_2^{2k} \left((-i)^{k}+ i^{k}\right)\\[4mm]
    &=& \displaystyle \frac{(\frac{1}{\sqrt{2}}\lambda_2)^{n-1}}{4}\left( (1+1)^{n-1}+(1+(-1))^{n-1}\right) + \frac{(\frac{1}{\sqrt{2}}\lambda_1)^{n-1}}{4}\left( (1+(-1))^{n-1}+(1+1)^{n-1}\right) \\[4mm]
    &=& \displaystyle \frac{(\sqrt{2}\lambda_1)^{n-1}+(\sqrt{2}\lambda_2)^{n-1}}{4},   
\end{array}    
\end{equation}
\begin{equation}\label{eq:Coeff3case2}
\begin{array}{rcl}
   &&\displaystyle\sum_{\mathbf{v}\in \mathbb{Z}_2^n, v_n=1} c_\mathbf{v} \frac{(-i)^{|\mathbf{v}|}+ i^{|\mathbf{v}|}}{2}\\[4mm] &=&\displaystyle\frac{1}{4}\sum_{\mathbf{v}\in \mathbb{Z}_2^n, v_n=1}  \left((\frac{1}{\sqrt{2}}\lambda_2)^{n-1} \lambda_1^{2|\mathbf{v}|}+(\frac{1}{\sqrt{2}}\lambda_1)^{n-1} \lambda_2^{2|\mathbf{v}|}\right)\left((-i)^{|\mathbf{v}|}+ i^{|\mathbf{v}|}\right)\\[4mm]
   &=& \displaystyle \frac{1}{4}\sum_{\mathbf{v}\in \mathbb{Z}_2^n, v_n=1, |\mathbf{v}|=k}    \left((\frac{1}{\sqrt{2}}\lambda_2)^{n-1} \lambda_1^{2k}+(\frac{1}{\sqrt{2}}\lambda_1)^{n-1} \lambda_2^{2k}\right)\left((-i)^{k}+ i^{k}\right)\\[4mm]
   &=& \displaystyle \frac{1}{4}\sum_{k=1}^{n}  \binom{n-1}{k-1}  \left((\frac{1}{\sqrt{2}}\lambda_2)^{n-1} \lambda_1^{2k}+(\frac{1}{\sqrt{2}}\lambda_1)^{n-1} \lambda_2^{2k}\right)\left((-i)^{k}+ i^{k}\right)\\[4mm]
   &=& \displaystyle \frac{1}{4}\sum_{\ell=0}^{n-1}  \binom{n-1}{\ell}  \left((\frac{1}{\sqrt{2}}\lambda_2)^{n-1} \lambda_1^{2\ell} i+(\frac{1}{\sqrt{2}}\lambda_1)^{n-1} \lambda_2^{2\ell} (-i) \right)\left((-i)^{\ell+1}+ i^{\ell+1}\right)\\[4mm]
    &=& \displaystyle \frac{i(\frac{1}{\sqrt{2}}\lambda_2)^{n-1}}{4}\sum_{\ell=0}^{n-1}  \binom{n-1}{\ell}    \lambda_1^{2\ell} \left((-i)^{\ell+1}+ i^{\ell+1}\right)+ \displaystyle \frac{-i(\frac{1}{\sqrt{2}}\lambda_1)^{n-1}}{4}\sum_{\ell=0}^{n-1}  \binom{n-1}{\ell}    \lambda_2^{2\ell} \left((-i)^{\ell+1}+ i^{\ell+1}\right)\\[4mm]
    &=& \displaystyle \frac{(\frac{1}{\sqrt{2}}\lambda_2)^{n-1}}{4}\left( (1+1)^{n-1}-(1+(-1))^{n-1}\right) + \frac{(\frac{1}{\sqrt{2}}\lambda_1)^{n-1}}{4}\left( -(1+(-1))^{n-1}+(1+1)^{n-1}\right) \\[4mm]
    &=& \displaystyle \frac{(\sqrt{2}\lambda_2)^{n-1}+(\sqrt{2}\lambda_1)^{n-1}}{4}.  
\end{array}
\end{equation}
So we have 
$$ \mathbf{I}^{(k)}_{n}(\theta) =\displaystyle \frac{(\sqrt{2})^{n-1}\left(\lambda_1 ^{n-1} +\lambda_2 ^{n-1}\right)}{4} \frac{\theta\gamma_k(\theta)+\theta  P_k(\theta)}{2^{k-1}} . $$
Note that 
$$
\lambda_1^{n-1}+\lambda_2^{n-1}=
\begin{cases}
    2(-1)^m=2(-1)^ {\lfloor \frac{n}{4}\rfloor}, & n=4m+1,\\
     \sqrt{2}(-1)^m=\sqrt{2}(-1)^ {\lfloor \frac{n}{4}\rfloor}, & n=4m+2,\\
         0 & n=4m+3,\\
              \sqrt{2}(-1)^{m+1}=\sqrt{2}(-1)^ {\lfloor \frac{n}{4}\rfloor}, & n=4m+4.
\end{cases}
$$
So 
$$
\displaystyle \frac{(\sqrt{2})^{n-1}\left(\lambda_1 ^{n-1} +\lambda_2 ^{n-1}\right)}{4}=
\begin{cases}
     (\sqrt{2})^{n-3}(-1)^ {\lfloor \frac{n}{4}\rfloor}, & n=4m+1,\\
    (\sqrt{2})^{n-4}(-1)^ {\lfloor \frac{n}{4}\rfloor}, & n=4m+2,\\
         0 & n=4m+3,\\
               (\sqrt{2})^{n-4}(-1)^ {\lfloor \frac{n}{4}\rfloor}, & n=4m+4.
\end{cases}
$$

As $\mathbf{I}^{(k)}_{n}(\theta) =0$ in the setting $n\equiv 3 \mod 4, $ we change the $A^{(n-1)}$'s observables   as follows  
\begin{eqnarray}\label{eq:casemod3}
 M^{(n-1)}_0 =-\theta\sigma_{2}, \ \
  M^{(n-1)}_1 = \theta\sigma_{1}.
\end{eqnarray} 
while the other are the same. For any vector  $\mathbf{v}=(v_1,v_2,\cdots,v_{n})\in \mathbb{Z}_2^{n},$   we redefine 
$\Tilde{\mathbf{v}}=(\Tilde{v}_1,\Tilde{v}_2,\cdots,\Tilde{v}_{n}):=(v_1+1,v_2+1,\cdots,v_{n-2}+1,\Tilde{v}_{n-1}, v_n+1)\in \{1,2\}^n$, when $v_{n-1}=0$ then $\Tilde{v}_{n-1}=2$ and when $v_{n-1}=1$ then $\Tilde{v}_{n-1}=1$. Note that 
$$ \left(\sigma_{\Tilde{v}_1} \otimes \sigma_{\Tilde{v}_2}\otimes \cdots \otimes \sigma_{\Tilde{v}_n}\right) |\mathrm{GHZ}_n\rangle =\begin{cases}
    \frac{1}{\sqrt{2}}( (i)^{|\mathbf{v}|+1} |11\cdots 1\rangle+(-i)^{|\mathbf{v}|+1}|00\cdots 0\rangle), \text{ if } v_{n-1}=0 \text{ and }v_n=0,\\[2mm]\frac{1}{\sqrt{2}}( (i)^{|\mathbf{v}|-1} |11\cdots 1\rangle+(-i)^{|\mathbf{v}|-1}|00\cdots 0\rangle),  \text{ if } v_{n-1}=1 \text{ 
and }v_n=0, \\[2mm]\frac{1}{\sqrt{2}}( (i)^{|\mathbf{v}|+1} |11\cdots 1\rangle+(-i)^{|\mathbf{v}|+1}|00\cdots 0\rangle),  \text{  if  } v_{n-1}=0 \text{ 
 and  }v_n=1,\\[2mm]\frac{1}{\sqrt{2}}( (i)^{|\mathbf{v}|-1} |11\cdots 1\rangle+(-i)^{|\mathbf{v}|-1}|00\cdots 0\rangle),  \text{  if  } v_{n-1}=1 \text{ 
 and  }v_n=1.
\end{cases}.$$

Therefore, 
$$\mathrm{Tr}[\rho^{(1) } \left(\sigma_{\Tilde{v}_1} \otimes \sigma_{\Tilde{v}_2}\otimes \cdots \otimes \sigma_{\Tilde{v}_n}\right) ]= \begin{cases}
    \frac{(-i)^{|\mathbf{v}|+1}+ i^{|\mathbf{v}|+1}}{2},  & \text{if } v_{n-1}=0 \text{ 
and }v_n=0, \\[2mm]
    \frac{(-i)^{|\mathbf{v}|-1}+ i^{|\mathbf{v}|-1}}{2},  &\text{ if } v_{n-1}=1 \text{ 
and }v_n=0,\\[2mm]
    \frac{(-i)^{|\mathbf{v}|+1}+ i^{|\mathbf{v}|+1}}{2},   & \text{if } v_{n-1}=0 \text{ 
and }v_n=1,\\[2mm]
    \frac{(-i)^{|\mathbf{v}|-1}+ i^{|\mathbf{v}|-1}}{2},  &\text{if } v_{n-1}=1 \text{ 
and }v_n=1.
\end{cases}$$ 
In this setting,  using \eqref{eq:recursive3cases} and repeating this recursive equation $(k-1)$ times,  one has that
  \begin{equation}\label{mkkk}
  \mathrm{Tr}[\rho^{(k)} \left(\sigma_{\Tilde{v}_1} \otimes \sigma_{\Tilde{v}_2}\otimes \cdots \otimes \sigma_{\Tilde{v}_n}\right)]= \begin{cases} 
  \frac{P_k(\theta)}{2^{k-1}}\frac{(-i)^{|\mathbf{v}|+1}+ i^{|\mathbf{v}|+1}}{2},  &\text{if } v_{n-1}=0 \text{ 
and }v_n=0,\\[2mm]\frac{P_k(\theta)}{2^{k-1}}\frac{(-i)^{|\mathbf{v}|-1} i^{|\mathbf{v}|-1}}{2},  &\text{if } v_{n-1}=1 \text{ 
and }v_n=0,\\[2mm]
  \frac{1}{2^{k-1}}
   \frac{(-i)^{|\mathbf{v}|+1}+ i^{|\mathbf{v}|+1}}{2},  &\text{if } v_{n-1}=0 \text{ 
and }v_n=1,\\[2mm]
  \frac{1}{2^{k-1}}
   \frac{(-i)^{|\mathbf{v}|-1}+ i^{|\mathbf{v}|-1}}{2},  &\text{if } v_{n-1}=1 \text{ 
and }v_n=1.
\end{cases}
\end{equation}
Now, using \eqref{mkkk}, we have \begin{equation}\label{eq:Mainmod3cases} 
    \mathrm{Tr}[\rho^{(k)}(M^{(1)}_{v_1}M^{(2)}_{v_2}\cdots M^{(n,k)}_{v_{n} })]=\begin{cases} -\theta  \mathrm{Tr}[\rho^{(k)}\left(\sigma_{\Tilde{v}_1} \otimes \sigma_{\Tilde{v}_2}\otimes \cdots \otimes \sigma_{\Tilde{v}_n}\right)],  &\text{if } v_{n-1}=0, v_{n}=0,\\[2mm]
    \theta  \mathrm{Tr}[\rho^{(k)}\left(\sigma_{\Tilde{v}_1} \otimes \sigma_{\Tilde{v}_2}\otimes \cdots \otimes \sigma_{\Tilde{v}_n}\right)],  &\text{if } v_{n-1}=1, v_{n}=0,\\[2mm]
   -\theta  \gamma_k(\theta) \mathrm{Tr}[\rho^{(k)}\left(\sigma_{\Tilde{v}_1} \otimes \sigma_{\Tilde{v}_2}\otimes \cdots \otimes \sigma_{\Tilde{v}_n}\right)],  &\text{if }  v_{n-1}=0, v_{n}=1,\\[2mm]
   \theta  \gamma_k(\theta) \mathrm{Tr}[\rho^{(k)}\left(\sigma_{\Tilde{v}_1} \otimes \sigma_{\Tilde{v}_2}\otimes \cdots \otimes \sigma_{\Tilde{v}_n}\right)],  &\text{if }  v_{n-1}=1, v_{n}=1. 
\end{cases}
\end{equation}

 Therefore,  by Eqs. \eqref{eq:Mermin_P}, \eqref{mkkk}, and \eqref{eq:Mainmod3cases}, the Mermin value $\mathbf{I}^{(k)}_{n}(\theta)$ of $\rho^{(k)}$ can be written as  
\begin{equation}\label{eq:Mermin_mod3}
\begin{array}{rcl}
 \displaystyle \mathbf{I}^{(k)}_{n}(\theta) &=& \displaystyle\left(-\sum_{\mathbf{v}\in \mathbb{Z}_2^n, v_{n-1}=0,v_n=0} c_\mathbf{v} \frac{(-i)^{|\mathbf{v}|+1}+ i^{|\mathbf{v}|+1}}{2}  -\sum_{\mathbf{v}\in \mathbb{Z}_2^n, v_{n-1}=1,v_n=0} c_\mathbf{v} \frac{(-i)^{|\mathbf{v}|-1}+ i^{|\mathbf{v}|-1}}{2}\right)\frac{\theta P_k(\theta)}{2^{k-1}}\\ [5mm]
 & &+ \displaystyle\left(-\sum_{\mathbf{v}\in \mathbb{Z}_2^n, v_{n-1}=0, v_n=1} c_\mathbf{v} \frac{(-i)^{|\mathbf{v}|+1}+ i^{|\mathbf{v}|+1}}{2} +\sum_{\mathbf{v}\in \mathbb{Z}_2^n, v_{n-1}=1, v_n=1} c_\mathbf{v} \frac{(-i)^{|\mathbf{v}|-1}+ i^{|\mathbf{v}|-1}}{2} \right)\frac{\theta  \gamma_k(\theta)}{2^{k-1}}.  
\end{array}    
\end{equation}
Similar with Eq. \eqref{eq:Coeff3case1} and \eqref{eq:Coeff3case2}, one finds that 
$$\begin{array}{c}
\displaystyle \sum_{\mathbf{v}\in \mathbb{Z}_2^n, v_{n-1}=0,v_n=0} c_\mathbf{v} \frac{(-i)^{|\mathbf{v}|+1}+ i^{|\mathbf{v}|+1}}{2}=\displaystyle \frac{(\sqrt{2})^{n-1}(i\lambda_1^{n-1}+(-i)\lambda_2^{n-1})}{8}=\displaystyle \sum_{\mathbf{v}\in \mathbb{Z}_2^n, v_{n-1}=0, v_n=1} c_\mathbf{v} \frac{(-i)^{|\mathbf{v}|+1}+ i^{|\mathbf{v}|+1}}{2}, \\[4mm]
 \displaystyle \sum_{\mathbf{v}\in \mathbb{Z}_2^n, v_{n-1}=1,v_n=0} c_\mathbf{v} \frac{(-i)^{|\mathbf{v}|-1}+ i^{|\mathbf{v}|-1}}{2}=\displaystyle \frac{(\sqrt{2})^{n-1}((-i)\lambda_1^{n-1}+i\lambda_2^{n-1})}{8}=\displaystyle \sum_{\mathbf{v}\in \mathbb{Z}_2^n, v_{n-1}=1, v_n=1} c_\mathbf{v} \frac{(-i)^{|\mathbf{v}|-1}+ i^{|\mathbf{v}|-1}}{2}.
 \end{array}$$
Therefore, both coefficients in Eq.\eqref{eq:Mermin_mod3} before the terms  $\frac{\theta P_k(\theta)}{2^{k-1}}$ and $\frac{\theta \gamma_k(\theta)}{2^{k-1}}$  are 
$$\displaystyle \frac{(\sqrt{2})^{n-1}((-i)\lambda_1^{n-1}+(i)\lambda_2^{n-1})}{4}=(\sqrt{2})^{n-3}(-1)^ {\lfloor \frac{n}{4}\rfloor}. $$
So we have 
$$ \mathbf{I}^{(k)}_{n}(\theta) =\displaystyle \displaystyle  (\sqrt{2})^{n-3}(-1)^ {\lfloor \frac{n}{4}\rfloor} \frac{\theta\gamma_k(\theta)+\theta  P_k(\theta)}{2^{k-1}}. $$

To sum up, we have 
$$ \mathbf{I}^{(k)}_{n}(\theta) =\displaystyle N_n\frac{\theta\gamma_k(\theta)+\theta  P_k(\theta)}{2^{k-1}}, $$
where \begin{equation}
    N_n=\begin{cases}
        (\sqrt{2})^{n-3}(-1)^ {\lfloor \frac{n}{4}\rfloor},  &\text{if } n \equiv 1 \text{ or }   3  \mod 4,\\[2mm](\sqrt{2})^{n-4}(-1)^ {\lfloor \frac{n}{4}\rfloor},  &\text{if } n \equiv 0 \text{ or }   2  \mod 4.
    \end{cases}
\end{equation}
\section{THE PROOF OF LEMMA \ref{Lemma}}\label{ap5}
 For any $k\in\{1,2,\cdots, K\}$, we need to find appropriate $\gamma_k(\theta)\in(0,1)$ and $\theta\in(-1,1)$ such that $f_k(\theta)=N\frac{\theta \gamma_k(\theta)+\theta P_k(\theta)}{2^{k-1}}>1$.  In the following $\theta$ is chosen according to the sign of $N$  such that $\theta N>0.$
 Since $$f_k(\theta)=N\frac{\theta \gamma_k(\theta)+\theta P_k(\theta)}{2^{k-1}}>1\Leftrightarrow \gamma_k(\theta)>\frac{2^{k-1}}{\theta N}-P_k(\theta).$$
 Therefore, for $\forall\epsilon>0$, we construct a specific sequences $\{\gamma_k(\theta)\}_{k=1}^K$. If $N>0$,  for $\theta\in(0,1)$, we define 
\begin{equation}\label{}
  \gamma_{k}(\theta)=\left\{
                       \begin{array}{ll}
                         (1+\epsilon)(\frac{1}{\theta N}-1), & \hbox{$k=1$;} \\[3mm]
                     (1+\epsilon)(\frac{2^{k-1}}{\theta N}-P_k(\theta)) , & \hbox{$0\leq\gamma_{k-1}(\theta)\leq1$ ;} \\[3mm]
                         \infty, & \hbox{otherwise.}
                       \end{array}
                     \right.
\end{equation}

Then we can get
\begin{equation}
  \frac{\gamma_{k}(\theta)}{\gamma_{k-1}(\theta)}>2\Leftrightarrow \frac{P_{k-1}(\theta)}{P_k(\theta)}>\frac{1}{2}\Leftrightarrow 0<\gamma_{k-1}(\theta)\leq1.
\end{equation} 
Here $\gamma_{1}(\theta)=(1+\epsilon)(\frac{1}{\theta N}-1)$, $\lim\limits_{\theta\rightarrow \frac{1}{N}}\gamma_{1}(\theta)=0$.
 By the induction, we can suppose there exists a $\delta_{k-1}$ such that for each $\theta$ on the interval $ (\frac{1}{N}-\delta_{k-1},\frac{1}{N})\cup(\frac{1}{N},\frac{1}{N}+\delta_{k-1})$(if $N=1$, we take the interval $(1-\delta_{k-1},1))$, all $\gamma_{i}(\theta)\in(0,1),$  and $\lim\limits_{\theta\rightarrow \frac{1}{N}}\gamma_{i}(\theta)=0$  for $i=1,2,\cdots ,k-1.$
 Then according to the definition of $\gamma_{k}(\theta)$, we will have $$\lim\limits_{\theta\rightarrow \frac{1}{N}}\gamma_{k}(\theta)=\lim\limits_{\theta\rightarrow \frac{1}{N}}  (1+\epsilon)(\frac{2^{k-1}}{\theta N}-P_k(\theta))=0,$$
here we use the limit $\lim\limits_{\theta\rightarrow \frac{1}{N}} P_k(\theta)=2^{k-1} $.   So  $\forall K\in\mathbb{N},$ we can find a $\delta_{K}>0$ such that $0<\gamma_{1}(\theta)<\gamma_{2}(\theta)<\cdots<\gamma_{K}(\theta)<1$ and $f_k(\theta)>1$ for all $ \theta\in(\frac{1}{N}-\delta_{K},\frac{1}{N})\cup(\frac{1}{N},\frac{1}{N}+\delta_{K})$ and $k\in\{1,2,\cdots,K\}$. If $N<0$, we then let $\theta\in(-1,0)$, which leads to the same conclusion. 
\section{THE CALCULATION OF $\mathbf{I}^{(k)}_{n}(\theta)$ IN DOUBLE-CHAIN SCENARIO }\label{ap6}
In this measurement strategy,  $A^{(1)}$'s observables are defined by
\begin{eqnarray}\label{eq:A1}
 M^{(1)}_0 =\theta\sigma_{1}, \ \
  M^{(1)}_1 = \theta\sigma_{2}.
\end{eqnarray}
$A^{(j)}$'s observables are defined by
\begin{eqnarray}\label{eq:A1}
 M^{(j)}_0 =\sigma_{1}, \ \
  M^{(j)}_1 = \sigma_{2}.
\end{eqnarray}
where $j=2,\cdots,n-2$.
 $A^{(n-1,k)}$'s $(1\leq k\leq K)$ observables are defined by
\begin{eqnarray}\label{eq:A1}
 M^{(n-1,k)}_0 =\sigma_{2}, \ \
  M^{(n-1,k)}_1 = \sigma_{2}.
\end{eqnarray}
$A^{(n,k)}$'s $(1\leq k\leq K)$ observables are defined by
\begin{eqnarray}\label{eq:A1}
 M^{(n,k)}_0 =\sigma_{1}, \ \
  M^{(n,k)}_1 =  {\gamma}_{k}(\theta)\sigma_{2},\ \  (0< {\gamma}_{k}(\theta)<1).
\end{eqnarray}  

 Let $\rho^{(k-1)}$ shared by $A^{(1)}$,  $A^{(2)}$, $\cdots$,  $A^{(n)}$, and $A^{(n,k-1)}$ prior to $A^{(n,k-1)}$'s measurements and with an initial state:  $\rho^{(1)}=\left|\mathrm{GHZ}_n\right\rangle\left\langle\mathrm{GHZ}_n\right|$, 
$\left|\mathrm{GHZ}_n\right\rangle=\frac{1}{\sqrt{2}}(\left|00\cdots0\right\rangle+\left|11\cdots1\right\rangle)$. Using the L\"{u}ders rule, the state sent to $A^{(n,k)}$ is
 \begin{equation}\label{gm1}
\begin{array}{rcl}
  \rho^{(k)} &=&\frac{1}{4}\sum(\mathbb{I} ^{\otimes (n-2)}\otimes\sqrt{M^{(n-1,k-1)}_{a^{(n-1,k-1)}|x^{(n-1,k-1)}}}\otimes\sqrt{M^{(n,k-1)}_{a^{(n,k-1)}|x^{(n,k-1)}}})\rho^{(k-1)}\\[4mm]&&(\mathbb{I} ^{\otimes (n-2)}\otimes\sqrt{M^{(n-1,k-1)}_{a^{(n-1,k-1)}|x^{(n-1,k-1)}}}\otimes\sqrt{M^{(n,k-1)}_{a^{(n,k-1)}|x^{(n,k-1)}}}). \\[4mm]&& =\frac{2+\sqrt{1-\gamma_{k-1}^2(\theta)}}{8}\rho^{(k-1)}+\frac{2+\sqrt{1-\gamma_{k-1}^2(\theta)}}{8}(\mathbb{I} ^{\otimes (n-2)}\otimes\sigma_2\otimes\mathbb{I})\rho^{(k-1)}(\mathbb{I} ^{\otimes (n-2)}\otimes\sigma_2\otimes\mathbb{I})\\[3mm]
&& +\frac{1}{8}(\mathbb{I} ^{\otimes (n-1)}\otimes\sigma_1)\rho^{(k-1)}(\mathbb{I} ^{\otimes (n-1)}\otimes\sigma_1)+\frac{1}{8}(\mathbb{I} ^{\otimes (n-2)}\otimes\sigma_2\otimes\sigma_1)\rho^{(k-1)}(\mathbb{I} ^{\otimes (n-2)}\otimes\sigma_2\otimes\sigma_1)\\[3mm]
&& +\frac{1-\sqrt{1-\gamma_{k-1}^2(\theta)}}{8}(\mathbb{I} ^{\otimes (n-1)}\otimes\sigma_2)\rho^{(k-1)}(\mathbb{I} ^{\otimes (n-1)}\otimes\sigma_2)+\frac{1-\sqrt{1-\gamma_{k-1}^2(\theta)}}{8}(\mathbb{I} ^{\otimes (n-2)}\otimes\sigma_2\otimes\sigma_2)\rho^{(k-1)}(\mathbb{I} ^{\otimes (n-2)}\otimes\sigma_2\otimes\sigma_2),
\end{array}    
\end{equation}
where we use Eq. (\ref{g1}).

To calculate $\mathbf{I}^{(k)}_{n}(\theta)$, by definition
\begin{equation}\label{gm2}
    \mathbf{I}^{(k)}_{n}(\theta)=\sum_{\mathbf{v}\in \mathbb{Z}_2^n} c_\mathbf{v} \mathrm{Tr}[\rho^{(k)}(M_{v_1}^{(1)}M_{v_2}^{(2)}\cdots M_{v_n}^{(n,k)})],
\end{equation} 
we should   calculate out each term $\mathrm{Tr}[\rho^{(k)}(M_{v_1}^{(1)}M_{v_2}^{(2)}\cdots M_{v_n}^{(n,k)})].$ For any vector  $\mathbf{v}=(v_1,v_2,\cdots,v_{n})\in \mathbb{Z}_2^{n},$   we define 
$\Tilde{\mathbf{v}}=(\Tilde{v}_1,\Tilde{v}_2,\cdots,\Tilde{v}_{n}):=(v_1+1,v_2+1,\cdots,v_{n-2}+1,\Tilde{v}_{n-1}, v_n+1)\in \{1,2\}^n$, with $\Tilde{v}_{n-1}=2$.
Note that 
$$ \left(\sigma_{\Tilde{v}_1} \otimes \sigma_{\Tilde{v}_2}\otimes \cdots \otimes \sigma_{\Tilde{v}_n}\right) |\mathrm{GHZ}_n\rangle =\begin{cases}
    \frac{1}{\sqrt{2}}( i^{|\mathbf{v}|+1} |11\cdots 1\rangle+(-i)^{|\mathbf{v}|+1}|00\cdots 0\rangle), &\text{ if } v_{n-1}=0 \text{ and }v_n=0,\\[2mm]\frac{1}{\sqrt{2}}( i^{|\mathbf{v}|} |11\cdots 1\rangle+(-i)^{|\mathbf{v}|}|00\cdots 0\rangle),  &\text{ if } v_{n-1}=1 \text{ 
and }v_n=0, \\[2mm]\frac{1}{\sqrt{2}}( i^{|\mathbf{v}|+1} |11\cdots 1\rangle+(-i)^{|\mathbf{v}|+1}|00\cdots 0\rangle), & \text{  if  } v_{n-1}=0 \text{ 
 and  }v_n=1,\\[2mm]\frac{1}{\sqrt{2}}( i^{|\mathbf{v}|} |11\cdots 1\rangle+(-i)^{|\mathbf{v}|}|00\cdots 0\rangle), &\text{  if  } v_{n-1}=1 \text{ 
 and  }v_n=1.
\end{cases}$$

Therefore, 
$$\mathrm{Tr}[\rho^{(1) } \left(\sigma_{\Tilde{v}_1} \otimes \sigma_{\Tilde{v}_2}\otimes \cdots \otimes \sigma_{\Tilde{v}_n}\right) ]= \begin{cases}
    \frac{(-i)^{|\mathbf{v}|+1}+ i^{|\mathbf{v}|+1}}{2},  & \text{if } v_{n-1}=0 \text{ 
and }v_n=0, \\[2mm]
    \frac{(-i)^{|\mathbf{v}|}+ i^{|\mathbf{v}|}}{2},  &\text{if } v_{n-1}=1 \text{ 
and }v_n=0,\\[2mm]
    \frac{(-i)^{|\mathbf{v}|+1}+ i^{|\mathbf{v}|+1}}{2},   & \text{if } v_{n-1}=0 \text{ 
and }v_n=1,\\[2mm]
    \frac{(-i)^{|\mathbf{v}|}+ i^{|\mathbf{v}|}}{2},  &\text{if } v_{n-1}=1 \text{ 
and }v_n=1.
\end{cases}$$ 

For each $\mathbf{v}=(v_1,v_2,\cdots,v_{n})\in \mathbb{Z}_2^{n},$ by substituting the $\rho^{(k)}$ with Eq. \eqref{gm1}, one finds that
\begin{equation}\label{gm3}
    \mathrm{Tr}[\rho^{(k)} \left(\sigma_{\Tilde{v}_1} \otimes \sigma_{\Tilde{v}_2}\otimes \cdots \otimes \sigma_{\Tilde{v}_n}\right)]=\begin{cases}
    \frac{1+\sqrt{1-\gamma_{k-1}^2(\theta)}}{2}\mathrm{Tr}[\rho^{(k-1)} \left(\sigma_{\Tilde{v}_1} \otimes \sigma_{\Tilde{v}_2}\otimes \cdots \otimes \sigma_{\Tilde{v}_n}\right)],  &\text{if } v_{n}=0,\\[2mm]
    \frac{1}{2}
    \mathrm{Tr}[\rho^{(k-1)} \left(\sigma_{\Tilde{v}_1} \otimes \sigma_{\Tilde{v}_2}\otimes \cdots \otimes \sigma_{\Tilde{v}_n}\right)],  & \text{if } v_{n}=1.
   \end{cases}
\end{equation} 
Repeating this recursive equation $(k-1)$ times,  one has that  
\begin{equation}\label{gm4}
  \mathrm{Tr}[\rho^{(k)} \left(\sigma_{\Tilde{v}_1} \otimes \sigma_{\Tilde{v}_2}\otimes \cdots \otimes \sigma_{\Tilde{v}_n}\right)]= \begin{cases} 
  \frac{P_k(\theta)}{2^{k-1}}\frac{(-i)^{|\mathbf{v}|+1}+ i^{|\mathbf{v}|+1}}{2},  &\text{if } v_{n-1}=0 \text{ and } v_{n}=0,\\[2mm]\frac{P_k(\theta)}{2^{k-1}}\frac{(-i)^{|\mathbf{v}|}+ i^{|\mathbf{v}|}}{2},  &\text{if } v_{n-1}=1 \text{ and } v_{n}=0,\\[2mm]
  \frac{1}{2^{k-1}}
   \frac{(-i)^{|\mathbf{v}|+1}+ i^{|\mathbf{v}|+1}}{2},  &\text{if } v_{n-1}=0 \text{ and } v_{n}=1,\\[2mm]
  \frac{1}{2^{k-1}}
   \frac{(-i)^{|\mathbf{v}|}+ i^{|\mathbf{v}|}}{2},  &\text{if } v_{n-1}=1 \text{ and } v_{n}=1,
\end{cases}
\end{equation}
One finds that 
\begin{equation}\label{gm5} 
    \mathrm{Tr}[\rho^{(k)}(M^{(1)}_{v_1}M^{(2)}_{v_2}\cdots M^{(n,k)}_{v_{n} })]=\begin{cases} \theta  \mathrm{Tr}[\rho^{(k)}\left(\sigma_{\Tilde{v}_1} \otimes \sigma_{\Tilde{v}_2}\otimes \cdots \otimes \sigma_{\Tilde{v}_n}\right)],  &\text{if } v_{n}=0,\\[2mm]
   \theta  \gamma_k(\theta) \mathrm{Tr}[\rho^{(k)}\left(\sigma_{\Tilde{v}_1} \otimes \sigma_{\Tilde{v}_2}\otimes \cdots \otimes \sigma_{\Tilde{v}_n}\right)],  &\text{if } v_{n}=1.
\end{cases}
\end{equation}
Therefore,  by Eqs. \eqref{gm2}, \eqref{gm4}, and \eqref{gm5}, the Mermin value $\mathbf{I}^{(k)}_{n}(\theta)$ of $\rho^{(k)}$ can be written as  
\begin{equation}\label{gm6}
\begin{array}{rcl}
 \displaystyle \mathbf{I}^{(k)}_{n}(\theta) &=& \displaystyle\left(\sum_{\mathbf{v}\in \mathbb{Z}_2^n, v_{n-1}=0,v_n=0} c_\mathbf{v} \frac{(-i)^{|\mathbf{v}|+1}+ i^{|\mathbf{v}|+1}}{2}  +\sum_{\mathbf{v}\in \mathbb{Z}_2^n, v_{n-1}=1,v_n=0} c_\mathbf{v} \frac{(-i)^{|\mathbf{v}|}+ i^{|\mathbf{v}|}}{2}\right)\frac{\theta P_k(\theta)}{2^{k-1}}\\ [5mm]
 & &+ \displaystyle\left(\sum_{\mathbf{v}\in \mathbb{Z}_2^n, v_{n-1}=0, v_n=1} c_\mathbf{v} \frac{(-i)^{|\mathbf{v}|+1}+ i^{|\mathbf{v}|+1}}{2} +\sum_{\mathbf{v}\in \mathbb{Z}_2^n, v_{n-1}=1, v_n=1} c_\mathbf{v} \frac{(-i)^{|\mathbf{v}|}+ i^{|\mathbf{v}|}}{2} \right)\frac{\theta  \gamma_k(\theta)}{2^{k-1}}.  
\end{array}    
\end{equation}
Now we calculate the two coefficients in the above equation ($\lambda_1^2=i$ and $\lambda_2^2=-i $)
\begin{equation}\label{gm7}
\begin{array}{rcl}
   &&\displaystyle\sum_{\mathbf{v}\in \mathbb{Z}_2^n, v_{n-1}=0,v_n=0} c_\mathbf{v} \frac{(-i)^{|\mathbf{v}|+1}+ i^{|\mathbf{v}|+1}}{2}\\[4mm] &=&\displaystyle\frac{1}{4}\sum_{\mathbf{v}\in \mathbb{Z}_2^n, v_{n-1}=0,v_n=0}  \left((\frac{1}{\sqrt{2}}\lambda_2)^{n-1} \lambda_1^{2|\mathbf{v}|}+(\frac{1}{\sqrt{2}}\lambda_1)^{n-1} \lambda_2^{2|\mathbf{v}|}\right)\left((-i)^{|\mathbf{v}|+1}+ i^{|\mathbf{v}|+1}\right)\\[4mm]
   &=& \displaystyle \frac{1}{4}\sum_{\mathbf{v}\in \mathbb{Z}_2^n, v_{n-1}=0,v_n=0, |\mathbf{v}|=k}    \left((\frac{1}{\sqrt{2}}\lambda_2)^{n-1} \lambda_1^{2k}+(\frac{1}{\sqrt{2}}\lambda_1)^{n-1} \lambda_2^{2k}\right)\left((-i)^{k+1}+ i^{k+1}\right)\\[4mm]
   &=& \displaystyle \frac{1}{4}\sum_{k=0}^{n-2}  \binom{n-2}{k}  \left((\frac{1}{\sqrt{2}}\lambda_2)^{n-1} \lambda_1^{2k}+(\frac{1}{\sqrt{2}}\lambda_1)^{n-1} \lambda_2^{2k}\right)\left((-i)^{k+1}+ i^{k+1}\right)\\[4mm]
    &=& \displaystyle \frac{(\frac{1}{\sqrt{2}}\lambda_2)^{n-1}}{4}\sum_{k=0}^{n-2}  \binom{n-2}{k}    \lambda_1^{2k} \left((-i)^{k+1}+ i^{k+1}\right)+ \displaystyle \frac{(\frac{1}{\sqrt{2}}\lambda_1)^{n-1}}{4}\sum_{k=0}^{n-2}  \binom{n-2}{k}    \lambda_2^{2k} \left((-i)^{k+1}+ i^{k+1}\right)\\[4mm]
    &=& \displaystyle \frac{(\frac{1}{\sqrt{2}}\lambda_2)^{n-1}}{4}\left((-i) (1+1)^{n-2}+i(1+(-1))^{n-2}\right) + \frac{(\frac{1}{\sqrt{2}}\lambda_1)^{n-2}}{4}\left( (-i)(1+(-1))^{n-2}+i(1+1)^{n-2}\right) \\[4mm]
    &=& \displaystyle \frac{(\sqrt{2})^{n-1}(i\lambda_1^{n-1}+(-i)\lambda_2^{n-1})}{8},   
\end{array}    
\end{equation}
\begin{equation}\label{gm8}
\begin{array}{rcl}
   &&\displaystyle\sum_{\mathbf{v}\in \mathbb{Z}_2^n,v_{n-1}=1, v_n=0} c_\mathbf{v} \frac{(-i)^{|\mathbf{v}|}+ i^{|\mathbf{v}|}}{2}\\[4mm] &=&\displaystyle\frac{1}{4}\sum_{\mathbf{v}\in \mathbb{Z}_2^n, v_{n-1}=1, v_n=0}  \left((\frac{1}{\sqrt{2}}\lambda_2)^{n-1} \lambda_1^{2|\mathbf{v}|}+(\frac{1}{\sqrt{2}}\lambda_1)^{n-1} \lambda_2^{2|\mathbf{v}|}\right)\left((-i)^{|\mathbf{v}|}+ i^{|\mathbf{v}|}\right)\\[4mm]
   &=& \displaystyle \frac{1}{4}\sum_{\mathbf{v}\in \mathbb{Z}_2^n, v_{n-1}=1, v_n=0, |\mathbf{v}|=k}    \left((\frac{1}{\sqrt{2}}\lambda_2)^{n-1} \lambda_1^{2k}+(\frac{1}{\sqrt{2}}\lambda_1)^{n-1} \lambda_2^{2k}\right)\left((-i)^{k}+ i^{k}\right)\\[4mm]
   &=& \displaystyle \frac{1}{4}\sum_{k=1}^{n-1}  \binom{n-2}{k-1}  \left((\frac{1}{\sqrt{2}}\lambda_2)^{n-1} \lambda_1^{2k}+(\frac{1}{\sqrt{2}}\lambda_1)^{n-1} \lambda_2^{2k}\right)\left((-i)^{k}+ i^{k}\right)\\[4mm]
   &=& \displaystyle \frac{1}{4}\sum_{\ell=0}^{n-2}  \binom{n-2}{\ell}  \left((\frac{1}{\sqrt{2}}\lambda_2)^{n-1} \lambda_1^{2\ell} i+(\frac{1}{\sqrt{2}}\lambda_1)^{n-1} \lambda_2^{2\ell} (-i) \right)\left((-i)^{\ell+1}+ i^{\ell+1}\right)\\[4mm]
    &=& \displaystyle \frac{(\frac{1}{\sqrt{2}}\lambda_2)^{n-2}}{4}\sum_{\ell=0}^{n-2}  \binom{n-2}{\ell}    \lambda_1^{2\ell} \left((-i)^{\ell}+ i^{\ell}\right)+ \displaystyle \frac{(\frac{1}{\sqrt{2}}\lambda_1)^{n-1}}{4}\sum_{\ell=0}^{n-2}  \binom{n-2}{\ell}    \lambda_2^{2\ell} \left((-i)^{\ell}+ i^{\ell}\right)\\[4mm]
    &=& \displaystyle \frac{(\frac{1}{\sqrt{2}}\lambda_2)^{n-1}}{4}\left( (1+1)^{n-2}-(1+(-1))^{n-2}\right) + \frac{(\frac{1}{\sqrt{2}}\lambda_1)^{n-1}}{4}\left( -(1+(-1))^{n-2}+(1+1)^{n-2}\right) \\[4mm]
    &=& \displaystyle \frac{(\sqrt{2})^{n-1}(\lambda_1^{n-1}+\lambda_2^{n-1})}{8}.  
\end{array}
\end{equation}
Similar with Eq. \eqref{gm7} and \eqref{gm8}, one finds that $$\displaystyle\sum_{\mathbf{v}\in \mathbb{Z}_2^n, v_{n-1}=0,v_n=0} c_\mathbf{v} \frac{(-i)^{|\mathbf{v}|+1}+ i^{|\mathbf{v}|+1}}{2}=\displaystyle \frac{(\sqrt{2})^{n-1}(i\lambda_1^{n-1}+(-i)\lambda_2^{n-1})}{8}=\displaystyle\sum_{\mathbf{v}\in \mathbb{Z}_2^n, v_{n-1}=0,v_n=1} c_\mathbf{v} \frac{(-i)^{|\mathbf{v}|+1}+ i^{|\mathbf{v}|+1}}{2},$$ and
$$\displaystyle\sum_{\mathbf{v}\in \mathbb{Z}_2^n, v_{n-1}=1,v_n=0} c_\mathbf{v} \frac{(-i)^{|\mathbf{v}|}+ i^{|\mathbf{v}|}}{2}=\displaystyle \frac{(\sqrt{2})^{n-1}(\lambda_1^{n-1}+\lambda_2^{n-1})}{8}=\displaystyle\sum_{\mathbf{v}\in \mathbb{Z}_2^n, v_{n-1}=1,v_n=1} c_\mathbf{v} \frac{(-i)^{|\mathbf{v}|}+ i^{|\mathbf{v}|}}{2}.$$
Therefore, both coefficients in Eq.\eqref{gm6} before the terms  $\frac{\theta P_k(\theta)}{2^{k-1}}$ and $\frac{\theta \gamma_k(\theta)}{2^{k-1}}$  are $$\displaystyle \frac{(\sqrt{2})^{n-1}((i+1)\lambda_1^{n-1}+(1-i)\lambda_2^{n-1})}{8}. $$
So we have 
$$ \mathbf{I}^{(k)}_{n}(\theta) =\displaystyle \frac{(\sqrt{2})^{n-1}\left((1+i)\lambda_1 ^{n-1} +(1-i)\lambda_2 ^{n-1}\right)}{8} \frac{\theta\gamma_k(\theta)+\theta  P_k(\theta)}{2^{k-1}} . $$
Note that 
$$
(i+1)\lambda_1^{n-1}+(i-1)\lambda_2^{n-1}=
\begin{cases}
    2(-1)^m=2(-1)^ {\lfloor \frac{n}{4}\rfloor}, & n=4m+1,\\
     0, & n=4m+2,\\
    (-2)(-1)^m=2(-1)^ {\lfloor \frac{n}{4}\rfloor+1} & n=4m+3,\\
              (-2\sqrt{2})(-1)^{m+1}=2\sqrt{2}(-1)^ {\lfloor \frac{n}{4}\rfloor+1}, & n=4m+4.
\end{cases}
$$
Then, it's easy to find that 
$$\displaystyle \frac{(\sqrt{2})^{n-1}\left((1+i)\lambda_1 ^{n-1} +(1-i)\lambda_2 ^{n-1}\right)}{8}=\begin{cases}
    (\sqrt{2})^{n-5}(-1)^ {\lfloor \frac{n}{4}\rfloor}, & n=4m+1,\\
     0, & n=4m+2,\\
    (\sqrt{2})^{n-5}(-1)^ {\lfloor \frac{n}{4}\rfloor+1} & n=4m+3,\\
              (\sqrt{2})^{n-4}(-1)^ {\lfloor \frac{n}{4}\rfloor+1}, & n=4m+4.
\end{cases} $$
As $\mathbf{I}^{(k)}_{n}(\theta) =0$ in the setting $n\equiv 2 \mod 4, $ we change the $A^{(n-1,k)}$'s $(1\leq k \leq K)$ observables   as follows  
\begin{eqnarray}\label{gm9}
 M^{(n-1,k)}_0 =-\sigma_{2}, \ \
  M^{(n-1,k)}_1 = \sigma_{2}.
\end{eqnarray}  
while the other are the same.
In this setting,  we also have equations \eqref{gm2}-\eqref{gm4} while Eq. \eqref{gm5} is replaced by the following equation
  \begin{equation}\label{gm10} 
    \mathrm{Tr}[\rho^{(k)}(M^{(1)}_{v_1}M^{(2)}_{v_2}\cdots M^{(n,k)}_{v_{n} })]=\begin{cases} -\theta  \mathrm{Tr}[\rho^{(k)}\left(\sigma_{\Tilde{v}_1} \otimes \sigma_{\Tilde{v}_2}\otimes \cdots \otimes \sigma_{\Tilde{v}_n}\right)],  &\text{if } v_{n-1}=0, v_{n}=0,\\[2mm]
   \theta  \mathrm{Tr}[\rho^{(k)}\left(\sigma_{\Tilde{v}_1} \otimes \sigma_{\Tilde{v}_2}\otimes \cdots \otimes \sigma_{\Tilde{v}_n}\right)],  &\text{if } v_{n-1}=1, v_{n}=0,\\[2mm]
   -\theta  \gamma_k(\theta) \mathrm{Tr}[\rho^{(k)}\left(\sigma_{\Tilde{v}_1} \otimes \sigma_{\Tilde{v}_2}\otimes \cdots \otimes \sigma_{\Tilde{v}_n}\right)],  &\text{if }  v_{n-1}=0, v_{n}=1,\\[2mm]
   \theta  \gamma_k(\theta) \mathrm{Tr}[\rho^{(k)}\left(\sigma_{\Tilde{v}_1} \otimes \sigma_{\Tilde{v}_2}\otimes \cdots \otimes \sigma_{\Tilde{v}_n}\right)],  &\text{if }  v_{n-1}=1, v_{n}=1. 
\end{cases}
\end{equation}

 Therefore,  by Eqs. \eqref{gm2}, \eqref{gm4}, and \eqref{gm10}, the Mermin value $\mathbf{I}^{(k)}_{n}(\theta)$ of $\rho^{(k)}$ can be written as  
\begin{equation}\label{gm11}
\begin{array}{rcl}
 \displaystyle \mathbf{I}^{(k)}_{n}(\theta) &=& \displaystyle\left(-\sum_{\mathbf{v}\in \mathbb{Z}_2^n, v_{n-1}=0,v_n=0} c_\mathbf{v} \frac{(-i)^{|\mathbf{v}|+1}+ i^{|\mathbf{v}|+1}}{2}  +\sum_{\mathbf{v}\in \mathbb{Z}_2^n, v_{n-1}=1,v_n=0} c_\mathbf{v} \frac{(-i)^{|\mathbf{v}|}+ i^{|\mathbf{v}|}}{2}\right)\frac{\theta P_k(\theta)}{2^{k-1}}\\ [5mm]
 & &+ \displaystyle\left(-\sum_{\mathbf{v}\in \mathbb{Z}_2^n, v_{n-1}=0, v_n=1} c_\mathbf{v} \frac{(-i)^{|\mathbf{v}|+1}+ i^{|\mathbf{v}|+1}}{2} +\sum_{\mathbf{v}\in \mathbb{Z}_2^n, v_{n-1}=1, v_n=1} c_\mathbf{v} \frac{(-i)^{|\mathbf{v}|}+ i^{|\mathbf{v}|}}{2} \right)\frac{\theta  \gamma_k(\theta)}{2^{k-1}}.  
\end{array}    
\end{equation}
Similar with Eq. \eqref{gm7} and \eqref{gm8}, one finds that both coefficients in Eq.\eqref{gm11} before the terms  $\frac{\theta P_k(\theta)}{2^{k-1}}$ and $\frac{\theta \gamma_k(\theta)}{2^{k-1}}$  are 
$$\displaystyle \frac{(\sqrt{2})^{n-1}((1-i)\lambda_1^{n-1}+(1+i)\lambda_2^{n-1})}{8}=\frac{(\sqrt{2})^{n}(-1)^ {\lfloor \frac{n}{4}\rfloor}}{4}= (\sqrt{2})^{n-4}(-1)^ {\lfloor \frac{n}{4}\rfloor} . $$
So we have 
$$ \mathbf{I}^{(k)}_{n}(\theta) =\displaystyle \displaystyle  (\sqrt{2})^{n-4}(-1)^ {\lfloor \frac{n}{4}\rfloor} \frac{\theta\gamma_k(\theta)+\theta  P_k(\theta)}{2^{k-1}}. $$

To sum up, we have 
$$ \mathbf{I}^{(k)}_{n}(\theta) =\displaystyle N_n\frac{\theta\gamma_k(\theta)+\theta  P_k(\theta)}{2^{k-1}}, $$
where \begin{equation}
    N_n=\begin{cases}
        \sqrt{2}^{n-4}(-1)^ {\lfloor \frac{n}{4}\rfloor+1},  &\text{if } n \equiv 0   \mod 4,\\[2mm]\sqrt{2}^{n-5}(-1)^ {\lfloor \frac{n}{4}\rfloor},  &\text{if } n \equiv 1   \mod 4,\\[2mm]\sqrt{2}^{n-4}(-1)^ {\lfloor \frac{n}{4}\rfloor},  &\text{if } n \equiv 2   \mod 4,\\[2mm]\sqrt{2}^{n-5}(-1)^ {\lfloor \frac{n}{4}\rfloor+1},  &\text{if } n \equiv 3   \mod 4.
    \end{cases}
\end{equation}

\twocolumngrid

\end{document}